\documentclass[12pt]{article}

\usepackage{mathrsfs}
\usepackage[T1]{fontenc}
\usepackage{mathpazo}
\usepackage{setspace}
\usepackage{amsfonts}
\usepackage{amssymb}
\usepackage{amsmath}
\usepackage{esint}                   
\usepackage{epsfig}
\usepackage{latexsym}
\usepackage{color}
\usepackage{graphicx}
\usepackage{hyperref}
\usepackage{nicefrac}
\usepackage[latin1]{inputenc}
\usepackage{pstricks}
\usepackage{slashed}
\usepackage{multirow}
\usepackage{ulem}
\usepackage{cancel}
\usepackage{comment}
\usepackage[nosort]{cite}
\usepackage{datetime}
\usepackage{tikz-cd}
\usepackage{float}

\usepackage{comment}
\usepackage{ytableau}
\usepackage{braket}

 \usetikzlibrary{decorations.pathreplacing}

\usepackage{pgfplots}
\usepackage{subcaption}

\usepackage{tikz}

 \usetikzlibrary{decorations.pathreplacing}

\def\hybrid{\topmargin -30pt    \oddsidemargin 0pt 
        \headheight 0pt \headsep 0pt
        \textwidth 6.25in       
        \textheight 9.5in       
        \marginparwidth .875in
        \parskip 5pt plus 1pt   \jot = 1.5ex}

\hybrid

\def\baselinestretch{1.2}

\catcode`\@=11

\def\marginnote#1{}
\newcount\hour
\newcount\minute
\newtoks\amorpm
\hour=\time\divide\hour by60
\minute=\time{\multiply\hour by60 \global\advance\minute by-\hour}
\edef\standardtime{{\ifnum\hour<12 \global\amorpm={am}%
        \else\global\amorpm={pm}\advance\hour by-12 \fi
        \ifnum\hour=0 \hour=12 \fi
        \number\hour:\ifnum\minute<10 0\fi\number\minute\the\amorpm}}
\edef\militarytime{\number\hour:\ifnum\minute<10 0\fi\number\minute}

\def\draftlabel#1{{\@bsphack\if@filesw {\let\thepage\relax
   \xdef\@gtempa{\write\@auxout{\string
      \newlabel{#1}{{\@currentlabel}{\thepage}}}}}\@gtempa
   \if@nobreak \ifvmode\nobreak\fi\fi\fi\@esphack}
        \gdef\@eqnlabel{#1}}
\def\@eqnlabel{}
\def\@vacuum{}
\def\draftmarginnote#1{\marginpar{\raggedright\scriptsize\tt#1}}

\def\draft{\oddsidemargin -.5truein
        \def\@oddfoot{\sl preliminary draft \hfil
        \rm\thepage\hfil\sl\today\quad\militarytime}
        \let\@evenfoot\@oddfoot \overfullrule 3pt
        \let\label=\draftlabel
        \let\marginnote=\draftmarginnote
   \def\@eqnnum{(\theequation)\rlap{\kern\marginparsep\tt\@eqnlabel}%
\global\let\@eqnlabel\@vacuum}  }

\def\draft2{
        \def\@oddfoot{\sl preliminary draft \hfil
        \rm\thepage\hfil\sl\today\quad\militarytime}
        \let\@evenfoot\@oddfoot \overfullrule 3pt
        \let\marginnote=\draftmarginnote
   \def\@eqnnum{(\theequation)\rlap{\kern\marginparsep\tt\@eqnlabel}%
\global\let\@eqnlabel\@vacuum}  }

\def\preprint{\twocolumn\sloppy\flushbottom\parindent 2em
        \leftmargini 2em\leftmarginv .5em\leftmarginvi .5em
        \oddsidemargin -.5in    \evensidemargin -.5in
        \columnsep .4in \footheight 0pt
        \textwidth 10.in        \topmargin  -.4in
        \headheight 12pt \topskip .4in
        \textheight 6.9in \footskip 0pt
        \def\@oddhead{\thepage\hfil\addtocounter{page}{1}\thepage}
        \let\@evenhead\@oddhead \def\@oddfoot{} \def\@evenfoot{} }

\def\numberbysection{\@addtoreset{equation}{section}
        \def\theequation{\thesection.\arabic{equation}}}

\def\underline#1{\relax\ifmmode\@@underline#1\else
        $\@@underline{\hbox{#1}}$\relax\fi}

\def\titlepage{\@restonecolfalse\if@twocolumn\@restonecoltrue\onecolumn
     \else \newpage \fi \thispagestyle{empty}\c@page\z@
        \def\thefootnote{\fnsymbol{footnote}} }

\def\endtitlepage{\if@restonecol\twocolumn \else \newpage \fi
        \def\thefootnote{\arabic{footnote}}
        \setcounter{footnote}{0}}  

\catcode`@=12
\relax

\def\figcap{\section*{Figure Captions\markboth
        {FIGURECAPTIONS}{FIGURECAPTIONS}}\list
        {Figure \arabic{enumi}:\hfill}{\settowidth\labelwidth{Figure
999:}
        \leftmargin\labelwidth
        \advance\leftmargin\labelsep\usecounter{enumi}}}
 \relax
\def\tablecap{\section*{Table Captions\markboth
        {TABLECAPTIONS}{TABLECAPTIONS}}\list
        {Table \arabic{enumi}:\hfill}{\settowidth\labelwidth{Table
999:}
        \leftmargin\labelwidth
        \advance\leftmargin\labelsep\usecounter{enumi}}}
 \relax
\def\reflist{\section*{References\markboth
        {REFLIST}{REFLIST}}\list
        {[\arabic{enumi}]\hfill}{\settowidth\labelwidth{[999]}
        \leftmargin\labelwidth
        \advance\leftmargin\labelsep\usecounter{enumi}}}
 \relax
\makeatletter
\newcounter{pubctr}
\def\publist{\@ifnextchar[{\@publist}{\@@publist}}
\def\@publist[#1]{\list
        {[\arabic{pubctr}]\hfill}{\settowidth\labelwidth{[999]}
        \leftmargin\labelwidth
        \advance\leftmargin\labelsep
        \@nmbrlisttrue\def\@listctr{pubctr}
        \setcounter{pubctr}{#1}\addtocounter{pubctr}{-1}}}
\def\@@publist{\list
        {[\arabic{pubctr}]\hfill}{\settowidth\labelwidth{[999]}
        \leftmargin\labelwidth
        \advance\leftmargin\labelsep
        \@nmbrlisttrue\def\@listctr{pubctr}}}
 \relax
\makeatother

\def\be{\begin{equation}}
\def\ee{\end{equation}}
\def\ba{\begin{eqnarray}}
\def\ea{\end{eqnarray}}

\def\del{\partial}

\def\bx {{\bf x}}
\def\bw {{\bf w}}

\def\k{\kappa}

\def\b{\beta}

\def\d{\delta}

\def\th{\theta}

\def\l{\lambda}

\def\no{\noindent}

\def\qq{\qquad}

\def\IR{\relax{\rm I\kern-.18em R}}

\def\bx {{\bf x}}
\def\bw {{\bf w}}

\def\ka {\kappa}

\def\inv{^{\raise.0ex\hbox{${\scriptscriptstyle -}$}\kern-.05em 1}}

\def \ha {{\frac{1}{2}}}

\def \ov {\over}

\def\diag{{\rm diag}}

\def\half{{\textstyle {1 \over 2}}}

\newcommand{\bb}{\hskip -0.1cm}
\newcommand{\hb}{\hskip -0.05cm}

\def\kk{{\bf k}}

\def\zz{{\bf z}}
\def\tr{\textrm{Tr}}

\usepackage{amsmath, amssymb, amsthm}
\usepackage{hyperref}
\usepackage{geometry}
\begin{document}


\renewcommand{\theequation}{\thesection.\arabic{equation}}
\csname @addtoreset\endcsname{equation}{section}

\begin{titlepage}
\begin{center}

\renewcommand*{\thefootnote}{\arabic{footnote}}

\phantom{xx}
\vskip 0.5in

{\large {\bf Thermodynamics of interacting magnetic chain molecules}} 

\vskip 0.5in

 {\bf Alexios P. Polychronakos}$^1$, \hskip .15cm
{\bf Konstantinos Sfetsos}$^{2}$

\vskip 0.15in

${}^1\!$ Physics Department, the City College of New York, NY 10031, USA \\
The Graduate Center of the CUNY, New York, NY 10016, USA\\
{\footnotesize{\tt apolychronakos@ccny.cuny.edu}}

\vskip 0.15in

${}^2\!$
Department of Nuclear and Particle Physics, \\
Faculty of Physics, National and Kapodistrian University of Athens, \\
Athens 15784, Greece\\
{\footnotesize{\tt ksfetsos@phys.uoa.gr}}

\vskip .3in
\today

\vskip .2in

\end{center}

\vskip .2in

\centerline{\bf Abstract}

\no
We study magnets composed of molecules with an arbitrary number $m$ of magnetically active atoms in a chain configuration interacting with
nearest-neighbor couplings. We derive the thermodynamics of a chain of spin-$j$ $SU(2)$ magnetic atoms using a mean-field approach and
determine the equilibrium states and stability conditions. The system exhibits a rich phase structure with a cascade of ferromagnetic phase
transitions, their number ranging from $m$, for maximal ferromagnets, to zero, for paramagnets, depending on the relative strengths of the couplings.
In the presence of impurities (modified couplings) at the ends of the chain, boundary states emerge that coexist with the paramagnetic phase
or the fully ferromagnetic phase.
For long chains, a continuum approximation can be taken that clarifies the structure of the phases. In the continuum description,
magnetized states at temperatures lower than the critical temperature at which they appear correspond to kink solitons.

\vskip .4in

\vfill

\end{titlepage}
\vfill
\eject



\def\baselinestretch{1.2}
\baselineskip 20 pt

\newcommand{\eqn}[1]{(\ref{#1})}

\tableofcontents

\section{Introduction}

$SU(N)$ magnetism is enjoying a revival, mostly because of its appearance in systems of ultracold molecules
\cite{Dud,Ghu,Gor,Zha,Mag,Cap,Mukherjee:2024ffz}. It has been theoretically studied in spin chains \cite{Aff,Pola} and
interacting molecules on lattice cites \cite{KT,BSL,RoLa,YSMOF,TK,Totsuka,TK2}, as well as in the presence of external $SU(N)$ magnetic fields \cite{DY,YM,HM}.

\no
In our recent work \cite{PSferro, PSlargeN, PSferroN, PShigher, PScubic, LPSadj} we derived the thermodynamics of $SU(N)$ ferromagnets in the mean field
approach and explored their phase structure,
which proved extremely rich and qualitatively different from that of ordinary $SU(2)$ ferromagnets. 
The magnets consisted of elementary components ("atoms") carrying internal degrees of freedom and interacting through two-body or higher-body interactions
invariant under global $SU(N)$ transformations of the atoms' degrees of freedom. Our study revealed the existence of several critical temperatures, 
metastability, hysteresis, and spontaneous breaking of the $SU(N)$ symmetry and of discrete symmetries.
The presence of three-body interactions introduces an additional relevant parameter that drives zero-temperature (quantum) phase transitions, while in the 
large-$N$ limit a triple critical point and new phases emerge.

\no
In a companion paper we extend our study to magnets with composite atoms ("molecules") consisting of several components (atoms)
with $SU(N)$ degrees of
freedom interacting through two-body terms \cite{mols}. Such magnets can have ferromagnetic, antiferromagnetic, or mixed properties, depending on the structure of their
coupling matrix. The equilibrium conditions of these magnets are derived and the simplest nontrivial case of two spin-half $SU(2)$ components, which is a generalization of the ordinary $SU(2)$ magnet, is worked out explicitly.

\no
In the present paper we extend the analysis to magnets consisting of macromolecules with many atoms in identical representations of $SU(2)$ coupled in a
chain-like fashion. The presence of this discrete internal dimension endows the system with additional structure and thermodynamic properties, involving
a cascade of critical temperatures and  several coexisting (metastable) states, including boundary and solitonic states.

\no
The organization of the paper is as follows:
In section \ref{model} we introduce the model of magnets with molecules having a general number of atoms in (generically different)
irreducible representations of $SU(N)$ and derive the equilibrium equations and stability conditions. In section \ref{chain} we study a chain of spin-$j$ $SU(2)$ atoms
and derive its thermodynamic equilibrium conditions and states, extended and boundary, and the critical temperatures. We also introduce a continuum version
of the model that serves to analyze states away from the critical temperatures, and identify soliton-like states at low temperatures.
Section \ref{conclusions} has our concluding remarks, and two appendices contain additional analysis and derivations.

\section{The multicomponent $SU(N)$ ferromagnet}\label{model}
 
 In this section we review the thermodynamics of systems of elementary magnets with many magnetic components interacting
 with two-body interactions.
 
 \subsection{Hamiltonian with two-body interactions}
 \label{tbH}
 
We consider $n \gg 1$ structurally equivalent molecules at fixed positions, each consisting of
$m$ atoms. The states in each atom transform under a global $SU(N)$ group in
irreducible representations (irreps) $r_s$, with $s=1,\dots,m$ labeling the atoms, the same for all molecules.
The Hilbert space ${\cal H}_I$ of each molecule $I=1,\dots,n$ is the direct product of the Hilbert spaces of its atoms ${\cal H}_{I,s} = {\cal H}_{s}$,
and the full Hilbert space ${\cal H}$ of the magnet is
\be
{\cal H} = \prod_{I=1}^n \prod_{s=1}^m \otimes {\cal H}_{I,s} = \Big(\prod_{s=1}^m \otimes {\cal H}_s\Big)^n \ , \quad \dim({\cal H}) = \prod_{s=1}^m \dim(r_s)^n\ .
\ee
Denoting $j_{s,i}^a$ the $SU(N)$ generators in irrep $r_s$ acting on the degrees of freedom $s$ of molecule $i$, the most general $SU(N)$ invariant Hamiltonian
with two-body interactions is
\be
H = \sum_{i,j=1}^n H_{ij} = \ha \sum_{i,j=1}^n \sum_{s,t=1}^m c_{st,ij}\sum_a j_{i,s}^a\, j_{j,t}^a\ ,
\label{Hgeneral}\ee
with $c_{st,ij}$ real coupling constants obeying
\be
c_{st,ij} = c_{ts,ji}\ ,\qq  c_{st,ii} = 0\ ,
\ee
the latter condition eliminating molecule self-interactions.

\subsection{The mean-field thermodynamic free energy} \label{thermofree}

In the thermodynamic limit of a large number of molecules $n$, a mean field analysis similar to that of \cite{PSferro, PShigher} shows that
the magnet can be described in terms of the global $SU(N)$ generators $J_s^a$ and effective coupling constants $C_{st}$
 \be
 \begin{split}
 &J_s^a = \sum_i j_{s,i}^a ~,~~~[J_s^a, J_t^b] = {\rm i} \,\delta_{st}\, \sum_c f^{abc} J_s^c \ ,
 \\
 &C_{st} = - \sum_{j} c_{st,ij}\ ,
 \label{Jsc}
 \end{split}
 \ee
 where assuming homogeneity of the magnet, the sum over molecules $j$ in \eqn{Jsc} is independent of the molecule $i$, and $C_{st}$ are globally defined quantities.
 The negative sign is included to make the ferromagetic case correspond to mostly positive $C_{st}$.
The resulting mean-field Hamiltonian is
 \be
 H = - \sum_{s,t=1}^m {C_{st}\over 2n} \sum_{a=1}^{N^2-1} J_s^a J_t^a
\label{Hcoupled}\ee
 and we refer the reader to \cite{PSferro, PShigher} for the details of its derivation. This Hamiltonian commutes with the total global $SU(N)$ generators
 \be J^a = \sum_{s=1}^m J_j^a
 \ee
 and, in principle, can be diagonalized within the space of irreps of $J^a$ contained in the direct product of the full set of individual irreps
 $(\prod_s \otimes r_s)^n$.
 
\no
The limit $n\gg 1$ is essentially a semiclassical limit and allows for further simplifications when we evaluate the partition function of the magnet.
We refer the reader to \cite{mols} for the derivations and simply state the result.

\no
The configuration of the system is parametrized by $m$ sets of $N$ non-negative distinct integers
$k_{s,1},\dots, k_{s,N}$, represented as the $N$-vectors $\kk_s$, $s=1,\dots,m$. Each $\kk_s$ parametrizes an irrep of $J_s^a$ embedded in the direct product
$(\otimes r_s)^n$ of the corresponding atoms in the Hilbert space, related to Young tableau row lengths as $\ell_i = k_i -k_N +i-N$ \cite{PScompo}.
The $k_i$ are, in principle, determined up to an overall additive constant (corresponding to the $U(1)$ charge of the irrep), but for irreps arising in the
decomposition of the direct product of $n$ irreps $r_s$ $(\otimes r_s)^n$, it is convenient to take $k_i$ to satisfy
 \be
 \sum_{i=1}^N k_i = n n_r +{N(N-1)\over 2}\ ,
 \ee
 with $n_r$ the number of boxes of irrep $r$, thus fixing their arbitrariness.
 
 \no
 The partition function is a sum over all allowed values of $\kk_s$ of a Boltzmann factor involving
 the quadratic Casimirs of irreps $\kk_s$, weighted by an appropriate multiplicity factor.
The $\kk_s$ dominating in the partition function are of order $n$. Setting $\kk_s = n {\bf x}_s$, turning the sums over $k_{s,i}$ into integrals over $x_{s,i}$, and
also introducing auxiliary $N$-vector variables $\bw_s$ (again, see \cite{PShigher,mols} for details), the partition function to leading order in $n$ is obtained as
\be
Z = \int \prod_{s=1}^m d{\bf x}_s\, d{\bf w}_s  \, e^{-n \beta F(\kk,\bw)}\ ,
\label{LZ}
\ee
with $d\bx = dx_1 \cdots dx_N$, $d\bw = dw_1 \cdots dw_N$ and
\be
F(\bx,\bw) = T\sum_s {\bx}_s \bb\cdot\bb{\bf w}_s - {1\over 2} \sum_{s,t} C_{st} {\bf x}_s \bb\cdot\bb {\bf x}_t
-T \sum_s \l_s ({\bw}_s)\ .
\label{Fn}\ee
The atom representations $r_s$ enter through the last, $\bw$-dependent term, given by
\be
\l_s = \ln \chi_s \ ,\qq \chi_s (\bw_s ) = \tr_{r_s} \diag \{e^{w_{s,1}} , \dots , e^{w_{s,N}} \}\ .
\label{zwlx}
\ee
That is, $\chi_s$ is the character of the irrep $r_s$ evaluated for a diagonal matrix with elements $e^{w_{s,i}}$.
The ${\bf x}_s$ are the order parameters of the magnet, representing $SU(N)$ magnetization per atom, $\bw_s$ are auxiliary variables that reproduce the entropy
and will be eliminated through their saddle point equations, and $F$ is the free energy per molecule.
 
 \subsection{Thermodynamic equilibrium equations and stability conditions}
 
 From the free energy $F(\bx,\bw)$ we can obtain the equilibrium equations and the free energy $F(\bx)$. For large $n$, the exponent in \eqn{LZ}
 is of order $n$ and the integrals can be evaluated by the saddle point method. The corresponding saddle point equations are
 \be
 \begin{split}
 & {\partial F \over \partial x_{s,i}} = 0 : \quad T\, {w}_{s,i} = \sum_t C_{st}\, {x}_{t,i} \ ,
 \\
 & {\partial F \over \partial w_{s,i}} = 0 : \quad  x_{s,i} = {\partial \l_s (\bw_s) \over \partial w_{s,i}} \ .
 \end{split}
 \label{saddle}
 \ee
Defining the $N$-vector function
\be   
{\bf B}_s = \{ B_{s,i} \}~,~~~  B_{s,i} (\bw_s )= {\partial \l_s (\bw_s) \over \partial {w_{s,i}} }\ ,
\label{Bs}\ee
the saddle point equations can be written in the more compact vector notation
 \be
T {\bf w}_s = \sum_t C_{st} {\bf x}_t ~,~~~~ \bx_s = {\bf B}_s (\bw_s) \ .
\label{xwB}\ee
Eliminating ${\bf w}_s$ gives the equilibrium equations for $\bx_s$
\be
\bx_s = {\bf B}_s \Big(\b\sum_t C_{st} {\bf x}_t \Big)
\label{sa}
\ee
and using the first equation in \eqn{xwB} to express $\bw_s$ in terms of $\bx_s$ and plugging in \eqn{Fn} yields the equilibrium free energy
\be
F_{eq} (\bx) = {1\over 2} \sum_{s,t} C_{st}\, {\bf x}_s \cdot{\bf x}_t
-T \sum_s \l_s \Big(\sum_t C_{st} \bx_t /T\Big)
\label{Fe}\ee
The derivation of the free energy \eqn{Fe} followed the path of solving the saddle point equations of $x_{s,i}$ for $w_{s,i}$ and is only valid for equilibrium
configurations satisfying \eqn{sa}. The off-equilibrium free energy, useful for deriving stability conditions, should, instead, involve eliminating
$\bw_s$ using their own saddle point equations.
%
The characters $\chi_s (\zz_s)$ are homogeneous polynomials in $z_{s,i}$ of degree $n_s$, the number of boxes in the
Young tableau of $r_s$. Summing the $w_{s,i}$ (second) saddle point equations in \eqn{saddle} over $i$ gives the constraints
\be
\sum_i x_{s,i} = n_s\ .
\label{cons}
\ee
Thus, solving these equations for the variables $w_{s,i}$ is possible only if the constraints are satisfied, and the
$w_{s,i}$ are determined up to an $i$-independent additive constant,
\be
w_{s,i}(\bx_s) \to w_{s,i} (\bx_s) +c_s (\bx_s)\ .
\label{wshift}
\ee
Calling $\bw_s (\bx_s)$ a solution of \eqn{saddle}, with $\bx_s$ satisfying the constraints \eqn{cons},
the free energy valid for off-equilibrium configurations is
\be
F (\bx) = T\sum_s {\bf x}_s \bb\cdot\bb\bw_s(\bx_s) - {1\over 2} \sum_{s,t} C_{st} {\bf x}_s \bb\cdot\bb {\bf x}_t
-T \sum_s \l_s (\bw_s(\bx_s))\ .
\label{Ft}\ee
For $\bx_s$ satisfying the constraints \eqn{cons}, $F$ is invariant under shifts \eqn{wshift}. In effect
\eqn{wshift} are gauge transformations and \eqn{cons} are their Gauss' law.

\no
The stability of equilibrium configurations is determined by the Hessian of the free energy \eqn{Ft} under variations of $x_{s,i}$ satisfying the
constraint $\sum_i \delta x_{s,i} =0$. Again, we refer the reader to \cite{PShigher,mols} and simply state the result.
Defining the $N$-dimensional matrix
\be
{\Lambda}_{s,ij} = {\partial^2 \lambda_s ({\bf w}_s) \over \partial_{w_{s,i}} \partial_{w_{s,j}}}\ ,
\ee
the stability criterion can be expressed as
\be
{\bf \Lambda} {\bf C}  \leqslant T\,  \mathbb{1} \quad \text{or}\quad \b{\bf \Lambda} {\bf C} \leqslant \mathbb{1}\ ,
\label{staba}
\ee
meaning that all the eigenvalues of the matrix $({\bf \Lambda C})_{si,tj} = \Lambda_{s,ij} C_{st}$ must be smaller than $T$.
The matrix $\bf \Lambda$ is positive definite \cite{LPSadj},
 but the matrix $C_{st}$ has no a priori positivity properties.


\section{Magnetic $SU(2)$-chain molecules}\label{chain}

We will now focus on "$SU(2)$ macromolecules," that is, molecules consisting of a large number of atoms carrying irreps of $SU(2)$. When the atom
irreps (spins) are all the same, and their interaction is nearest-neighbor and of constant strength, we will call this magnet an $SU(2)$ chain.

\no
We begin by reviewing the case of single-atom $SU(2)$ molecules with spin-$j$ states, which reproduces the usual ferromagnet.
Its thermodynamics can be obtained by setting $m=1$, eliminating the index $s$ in $x_{s,i},w_{s,i}$, and turning $C_{st}$ into a scalar $c$. 
A spin-$j$ representation corresponds to $2 j$ boxes in the Young Tableau, and the corresponding character is
\be
\label{xi1}
\chi_j = 
e^{j(w_1+w_2)} {\sinh {(j+1/2) (w_1-w_2)}\ov \sinh {(w_1-w_2)/2}}
\ee
and the constraint is $x_1+x_2=2 j$. The factor $e^{j(w_1+w_2)} $ is physically irrelevant (a $U(1)$ charge) and can be dropped. 
In the process we effectively shift $x_i\to x_i+j$, so that the new $x_{1,2}$ sum to zero. Denoting by $w=w_1-w_2$, the character becomes
\be
\label{xi2}
\chi_j =  {\sinh {(j+1/2) w}\ov \sinh {\half w}}\ .
\ee
The variables $x_1 = -x_2$ take values in the range $[-j,j]$, so 
we define $x_1= - x_2 = j x$ with $x\in [-1,1]$. Then, the free energy becomes
\be
F(x,w) = T j x w - c j^2  x^2 - T \ln {\sinh {(j+1/2) w}\ov \sinh {w/2}}\ .
\ee
The equilibrium equations for $x$ and $w$ are, respectively,
\be
 w = {2\ov T} j c x\ ,\qq   x = B_j(j w) \ ,
\ee
where 
\be
B_j(s) = {1+2 j\ov 2j} \coth  {1+2 j\ov 2j} s -{1\ov 2j} \coth {s\ov 2 j}\ ,
\ee
is the well known Brillouin function, which is the general function ${\bf B}_s$ \eqn{Bs} reduced to the case of a single $SU(2)$ atom. 
$B_j(s)$ is an odd  function in $s$, monotonically increasing from $-1$ to $1$. (It is concave, i.e  $B''_j(s)<0$, for $s>0$  and convex,  i.e  $B''_j(s)>0$ 
for $s<0$.)
Eliminating $w$ we obtain the transcendental equation 
\be
 x = B_j(2 c j^2 x/T)\ .
\ee
This classic result can also be  obtained  by elementary means in statistical mechanics using the mean field approximation. 
The solution of this equation is  $x=0$ for $T>T_C$ and $x\neq 0$ for $T<T_C$, where the critical Curie temperature is
\be
T_C = \frac{2\, j (j+1)}{3} c \ .
\ee
Note that the classical case of $j\to \infty$ is also attainable by first rescaling $c\to c/j^2$. Then 
\be
 x = L(2 c x/T)\ ,
\ee
where 
\be
L(s) = \coth  s -{1\ov s}\ ,
\ee
is the Langevin function, the large-spin limit of the Brillouin function. 

\subsection{$SU(2)$ chains of magnets}

For a system with molecules consisting of $m$ $SU(2)$ atoms,
in a treatment similar to the one around \eqn{xi1} and \eqn{xi2}, we define new variables satisfying the constraints
in $x_{s,i}$ and eliminating the arbitrariness in $w_{s,i}$
\be
x_{s,1} = - x_{s,2} =j_s x_s\ ,\qq w_{s,1}= -w_{s,2} ={w_s\ov 2}\ .
\ee
In the new variables, the free energy \eqn{Fn} becomes
\be
F = T \sum_{s=1}^m j_s x_s w_s - \sum_{s,t=1}^m C_{st} j_s j_t x_s x_t - T \sum_{s=1}^m \ln {\sinh (j_s+\ha) w_s\ov \sinh {\ha w_s}}\ .
\label{Fred}\ee
The equilibrium equations for $x_s$ and $w_s$ are, respectively,
\be
w_s = {2\ov T} \sum_{t=1}^m C_{st} j_t x_t\ , \qq x_s = {1\over j_s} {\partial \over \partial w_s}  \ln {\sinh (j_s+\ha) w_s\ov \sinh {\ha w_s}}
= B_{j_s} (j_s w_s)
\ee
and eliminating $w_s$,
\be
x_s = B_{j_s}\bigg({2j_s \ov T} \sum_{t=1}^m C_{st} j_t x_t\bigg)\ ,\qq s=1,2,\dots , m\ ,
\label{xsB}\ee
obtaining a system of $m$ coupled transcendental equations in the $x_s$. 
Calling  $G_j :=B_j^{-1}$ the inverse function of $B_j$ (also odd and monotonically increasing, but convex for $s>0$ and concave for $s<0$),
\eqn{xsB} can be written as
\be
\sum_{t=1}^m C_{st} j_t x_t = {T\over 2j_s} G_{j_s} (x_s )\ ,\qq s=1,2,\dots , m\ .
\label{xsG}\ee
In the limit of large spins $j_s\to \infty$, upon rescaling $C_{st}\to {C_{st}\ov j_s j_t}$, the equilibrium equations are
\be
 \sum_{t=1}^m C_{st} x_t = {T\over 2} G (x_s ) \ , \qq s=1,2,\dots , m\ ,
\ee
with $G(x)$ the inverse of the Langevin function $L(x)$.

\no
Finally, eliminating $w_s$ using its own equilibrium equation, that is, setting $w_s = G_{j_s} (x_s )/j_s$
and substituting into \eqn{Fred}, we obtain the off-equilibrium free energy in terms of $x_s$
\be
F = T \sum_{s=1}^m x_s G_{j_s} (x_s) - \sum_{s,t=1}^m C_{st} j_s j_t x_s x_t - T \sum_{s=1}^m \ln {\sinh [(1+{1\over 2j_s}) G_{j_s} (x_s)]\ov \sinh {[{1\ov 2j_s} G_{j_s}(x_s)}]}\ .
\label{FG}
\ee
Differentiating the above $F$ with respect to $x_s$ yields the equilibrium equation \eqn{xsG}.

\no
So far, the discussion has been general.
In the sequel, we will focus on molecules realizing $SU(2)$ spin chains, in which the  interactions $C_{st}$ are of nearest-neighbor form.

\subsection{Constant nearest-neighbor interactions}

We now specialize to the interesting case of $m$ atoms with nearest-neighbor interactions
\be
\label{CRS}
C_{rs} = {\hat c}_s \d_{sr}  + \half c_r \d_{s,r+1}  + \half c_s \d_{r,s+1} \  ,\qq s,r=1,\dots,m\ .
\ee
Then the equilibrium conditions \eqn{xsG} become the difference equations\footnote{
For $j_s=\ha$ this system is similar to the McCoy-Wu 
Ising model \cite{McCoy-Wu,McCoy} in the mean field approximation \cite{McCoyMean}. In this model the bonding of ferromagnetic layers,
corresponding to couplings ${\hat c}_s,c_s$ in our case, is via randomly distributed couplings. To the best of our knowledge the system 
\eqn{xsdis0} has not
been studied for general quantum spin $j_s$.
}
\be
\label{xsdis0}
2 j_s{\hat c}_s  x_s +j_{s-1} c_{s-1}  x_{s-1} + j_{s+1}c_s  x_{s+1} = {T\over j_s} G_{j_s} (x_s)\, , \quad s=1,\dots,m\, ,\quad  x_0 = x_{m+1} =0\ .
\ee
The situation further simplifies if we take all spins to be the same, and interactions to be homogeneous, that is, $j_s = j$, 
${\hat c}_s = c_0$, $c_s = c$. In this case, the equilibrium equations are
\be
\label{xsdis}
2 c_0 x_s + {c} (x_{s-1} + x_{s+1}) =  {T \over j^2 }  G_j (x_s )~,\quad \quad x_0 = x_{m+1} \equiv 0
\ee
and the corresponding free energy is
\be
F = T \sum_{s=1}^m x_s G_j (x_s) - j^2\sum_{s=1}^m (c_0 x_s^2 +c x_s x_{s-1}) 
- T \sum_{s=1}^m \ln {\sinh (1+{1\ov 2 j}) G_j (x_s)\ov \sinh {{1\ov 2j} G_j(x_s)}}\  .
\label{Fchain}
\ee
Note that the sign of $c$ can be flipped by redefining $x_s \to (-1)^s x_s$, so we can take $c> 0$ from now on (if $c=0$, the magnet becomes
a collection of $m$ noninteracting magnets). The sign of $c_0$,
however, cannot be redefined and remains relevant. The system is invariant under $x_s \to -x_s$ (spin flip) and $x_s \to x_{m+1-s}$ (inversion
of the chain).

\no
The system \eqn{xsdis} can, in principle, be solved by assuming an arbitrary value for $x_1$ and solving the $s=1$ equation for $x_2$, and iteratively for
the remaining $x_s$, and finally setting $x_{m+1}=0$, which fixes the allowed values of $x_1$ and corresponding solutions. In practice, this can only be
carried out numerically. In what follows, we will work in two special limits and will investigate the relevant phase transitions.

\subsection{States close to the singlet}

We start by considering states in the neighborhood of the singlet state $x_s =0$.
Assuming small magnetizations for all atoms, i.e., $x_s\ll 1$,  we expand \eqn{xsdis} as 
\be
2c_0 x_s + c x_{s+1} + c x_{s-1} - {3 T \over j(j+1)} x_s= a T x_s^3 + {\cal O}(x_s^5)\ ,
\label{xex}
\ee
where we defined the constant $a$
\be
\label{ggga}
a =  {9(2j^2+2j+1)  \over 10\, j(j+1)^3 } \ .
\ee
Keeping only the linear terms leads to a homogeneous equation with general real solution
\be
x_s = 2 A \sin(\theta s + \d) \ ,\qq \cos \theta = {3T\over 2cj(j+1)}-{c_0\over c}\ .
\label{xlin}
\ee
For
\be
\left|{3T\over 2cj(j+1)}-{c_0\over c}\right| < 1 \quad \Rightarrow\quad {2j(j+1)\over 3}(c_0 -c) < T < {2j(j+1)\over 3} (c_0 +c)\ ,
\label{critcon}
\ee
the parameter $\theta$ is real, and implementing the boundary conditions $x_0 = x_{m+1} = 0$
we obtain for the phase shift $\d=0$ ($\d=\pi$ corresponds to $A \to -A$) and the solutions
\be
\theta = {\pi k \over m+1} \ ,\qq x_s = 2 A \sin {\pi k s\over m+1} \ ,\qq k=1,\dots, m\ .
\label{xsm}
\ee
If \eqn{critcon} is not satisfied, $\theta$ has an imaginary part and \eqn{xlin} yields no nonzero solutions that satisfy the boundary conditions.
Therefore, solutions exist for the specific values of $\theta$, and through \eqn{xlin}, of $T$,
\footnote{\label{foot2}
For $m=2$ and $j=\half$ we obtain $T_1={2 j(j+1)\ov 3} (c_0 + c/2)$ and $T_2= {2 j(j+1)\ov 3} (c_0 - c/2)$, which coincide 
with $ T_\pm$ in eq. (3.29) of \cite{mols} upon setting in that work $c_a=c_b=c_0$ and $c \to c/2$.}
\be
T_k = {2\over 3}j(j+1) \Big( c_0 + c \cos{\pi k\over m+1}\Big) \ ,\qq  T_{k+1}<T_k\ ,
\label{Tk}
\ee
for all $k=1,\dots,m$ for which $T_k$ is positive (which implies \eqn{xsm} is satisfied). For $c>|c_0|$, we obtain a truncated set of critical temperatures.
For $c<|c_0|$, there will be either a full set of $m$ critical temperatures, for $c_0>0$,
or no critical temperatures, for $c_0 <0$.

\no
The existence of nonzero solutions for small $x_s$ signals the appearance of additional solutions of the equilibrium equation
close to the singlet $x_s=0$. So, the temperatures $T_k$ are critical temperatures at which additional nonsinglet solutions start to emerge.

\no
To find the actual form of $x_s$ and lift the linear degeneracy of the solutions (fix the arbitrary coefficient $A$ in \eqn{xsm})
at temperatures near a critical value $T_k$, higher order terms have to be included in \eqn{xex}. Focusing on temperatures close to $T_k$,
the form of the solutions will remain close to \eqn{xsm} but will develop a perturbation.
Retaining the cubic term in \eqn{xex}, we write $x_s$ as a sum of a solution of the linear equation at $T = T_k$ plus a vector $y_s$
\be
x_{k;s} = 2 A u_{k;s} + y_{s}\ ,\qq  u_{k;s} = \sin{\pi k s\over m+1}\ ,\qq  |y_s| \ll A \ll 1\ .
\label{xsnonlin}\ee
We may take the vector $y_s$ orthogonal to $u_{k;s}$, i.e.,
\be
  \sum_{s=1}^m u_{k;s} y_{s} = 0\ .
 \ee
Substituting in \eqn{xex}, omitting terms $\sim y_s/A$, and using the trigonometric identity
\be
u_{k;s}^3 = {3\over 4} u_{k,s} - {1\over 4} u_{3k;s}\ ,
\ee
we obtain
\be 
\bigg(2c_0 - {3 T \over j(j+1)}\bigg)  y_s + c (y_{s+1} + y_{s-1}) =  6 A \Big({a A^2} T + {T-T_k \over j(j+1)}\Big) u_{k;s}  
- 2 {a A^3} T u_{3k;s}\ .
\label{yu}
\ee
Contracting \eqn{yu} with $u_{k;s}$, shifting summation indices in the left hand side, using the expression
for $u_{k;s}$ (which makes the left hand side proportional to $ \sum_{s=1}^m u_{k;s} y_{s} $), and using the orthogonality of $u_{k;s}$ with $y_s$ and
$u_{3k;s}$ yields
\be
 6A \Big({a A^2} T+  {T-T_k \over j(j+1)}\Big) \sum_{s=1}^m u_{k;s}^2 = 0\ ,
\ee
giving, after using \eqn{ggga}, that
\be
A = {j+1\ov 3}   \sqrt{10\ov  2j^2 + 2j +1}\,  \sqrt{ T_k - T \ov T} \ .
\label{finA}
\ee
We see that the size of the magnetization $A$ has the typical $\sim\sqrt{T_\text{critical} - T}$ behavior of ferromagnetic phase transitions.
Finally, from the form of \eqn{yu} we see that $y_s$ will only contain the mode $u_{3k;s}$. Contracting \eqn{yu} with $u_{3k;s}$,
shifting summation indices, and using the explicit form of $u_{3k;s}$, we obtain
\be
y_{s} = {2 A\ov 3}  {T_k-T\ov T-T_{3 k}}  u_{3k;s}
=   {2\ov 9}   (j+1)  \sqrt{10\ov  2j^2 + 2j +1}\, \sqrt{ T_k - T \ov T}\, {T_k-T\ov T-T_{3 k}}  \sin{3\pi k s\over m+1}\ .
\ee
Thus, for $T$ very close to $T_k$
the magnetization is proportional to $u_{k;s}$, but to higher order it develops a magnetization mode proportional to $u_{3k;s}$.

\no
We note that the above results are exact in the temperature, but the assumption of small $x_s$ (small $A$ in \eqn{finA})
requires $T$ to be close to $T_k$. For larger $T_k-T$ we would need to keep higher order terms in $x_s$ in the expansion of \eqn{xsdis}, which would have
produced modes proportional to $u_{5k}, u_{7k}, \dots$ with diminishing coefficients.
Note also that the above results are valid up to "umklapp" effects, since
\be
u_{k;s} = \pm u_{2(m+1)\pm k;s} \ ,\qq u_{m+1;s} =0
\ee
and thus for $k=(m+1)/2$ the modes $u_{k;s}$ and $u_{3k;s}$ are not orthogonal. In these cases the
value of the amplitude $A$ differs from the one in \eqn{finA}, and no additional modes arise to cubic order in $x_s$. 
\no
To determine the physical relevance of these solutions we need to consider their stability. For configurations close to the singlet $\bx_s =0$,
the general equilibrium equation \eqn{saddle}, when expanded to first order near $\bx_s=0$ (that is, $\bw_s=0$), yields
\be
x_{s,i} =  \sum_j {\del^2 \l \over \del w_{s,i}\del w_{s,j}}\bigg |_{\bw_s=0} w_{s,j}+\dots  = \b\sum_{t,j} \Lambda_{s,ij} C_{st} x_{t,j} +\dots
= \b({\bf \Lambda C} \,\bx)_{s,i}+\dots\ .
\ee
The matrix $\b {\bf \Lambda C}$ that appears in the right hand side is precisely the matrix entering the stability condition \eqn{staba}. 
Therefore, the solution \eqn{xsm} of the linear part
of the equilibrium equation corresponds to a unit eigenvalue of the matrix $\b {\bf \Lambda} {\bf C}$. 
The full stability of the singlet is determined by the remaining eigenvalues of $\b {\bf \Lambda} {\bf C}$, which must all be less than 1. Solving the
eigenvalue equation
\be
{j(j+1)\ov 3 T} \big(2c_0  x_s + c x_{s-1} + c x_{s+1}) = \l x_s\ ,
\ee
we find the eigenvalues corresponding to eigenvectors $u_{k;s}$
\be
\lambda_k = {2j(j+1)\over 3T} \Big( c_0 + c \cos{\pi k\over m+1}\Big) = {T_k \over T} \ .
\label{lam}\ee
Clearly the greatest eigenvalue (for positive $c$) is $\lambda_1$. Stability of the singlet requires
\be
T > T_1 = {2\over 3}j(j+1) \Big( c_0 + c \cos{\pi \over m+1}\Big)\ .
\ee
Thus, the critical temperature below which the singlet destabilizes is $T_1$, and the state $x_{1;s}$ appears.
At the other temperatures $T_k < T_1$, new nonsinglet states start to appear, while the singlet remains unstable. Notably, for $c_0 < -c$,
no critical temperature emerges and the singlet is always stable. This is a purely paramagnetic phase.

\no
The $k=1$ state $x_{1;s}$ is stable for $T<T_1$, inheriting the stability of the just destabilized singlet mode. The states $x_{k;s}$ for $k>1$, however,
are unstable for $T \lesssim T_k$, since their stability matrix is perturbatively close to the one of the singlet at this temperature, which has $k-1$ unstable
(less than 1) eigenvalues. Thus, states $x_{k;s}$ for $k>1$ near their critical temperature $T_k$ are not physical. As $T$ further decreases, their
stability condition evolves away from the perturbative one and they can become stable.
The results of the next section show that, indeed, such a stabilization does occur.

\no
 Note that other states may appear at low temperatures with no counterpart 
in the linear approximation. Indeed, the exhaustive investigation of the $m=2$ case in \cite{mols} reveals that this is the case
at the full nonlinear level. At the temperature that the mode $k=2$ stabilizes ($T_c$ in the notation of \cite{mols}, see also footnote \ref{foot2}),
 two extra modes appear which nevertheless are unstable, see fig. 6 of \cite{mols}.

\subsection{The continuum limit}

\no
The previous analysis exhausts the states of the system close to the singlet and close to a critical temperature.
A different regime in which we can study the system that will allow us to examine configurations away from the singlet
is the limit $m \gg 1$ (but still $m$ much smaller than the number of molecules $n$) , and for excitations that vary on the scale of $m$, that is, where the configuration $x_s$ can be approximated by a continuous
function $x(s)$. Substituting $x(s)$ in \eqn{xsdis}, and using
\be
\label{derexpa}
x_{s\pm 1} = x(s) \pm x'(s) +\ha x''(s) \pm{1\over 6} x''' (s) + \dots \ ,
\ee
we obtain from \eqn{xsdis}, ignoring terms $O(x^{(4)})$ and writing simply $x$ for $x(s)$,
\be
c x'' +2(c_0 + c) x =  {T\over j^2} G_j (x)  \ .
\label{diffm1}\ee
The continuum system inherits the invariances of the discrete system $x(s) \to -x(s)$ and $x(s) \to x(m-s)$.

\no
The equation \eqn{diffm1} arises from the stationary points of the free energy of the system \eqn{Fchain}, which, in the continuum approximation,
becomes
\be
\label{Fcont}
F = j^2 \int_0^m ds \left[ {c\ov 2} x'^2 - V(x) \right]\ ,
\ee
where
\be
V(x) = (c_0 +c) x^2 -{T\over j^2}x\, G_j (x)+{T\over j^2} \ln {\sinh\hb{(2j+1)) G_j (x)\over 2j} \over (2j\hb+\hb 1)\sinh\hb{G_j (x)\over 2j}}\ ,
\ee
(we have subtracted the constant  $\displaystyle T \ln (2 j+1)/j^2$ from \eqn{Fchain} so that $V(0)=0$.)
The free energy has the form of the action of a fiducial particle with coordinate $x$ and mass $c$, with $s$ playing the role of time, in the potential $V(x)$.
Equation \eqn{diffm1} can be written as
\be
c x'' + 2(c_0 + c) x - {T\over j^2} G_j (x) = c x'' + {dV (x) \over dx} = 0\ .
\label{xmotion}\ee
Integrating once, we obtain the first-order equation
\be
{c\ov 2} x'^2 + V(x) = E\ ,
\label{anharmonic}\ee
the integration constant $E$ playing the role of fiducial energy.
Note that, since the free energy must be minimized, the potential $V(x)$ for stable constant configurations with $x'(s)=0$ should be at a maximum.

\no
In general, the form of $V(x)$ cannot be found analytically, except for $j=\ha,1,{3\over 2},2$, where deriving $G_j(x)$ reduces to finding the roots of an
algebraic equation of degree $2j$. The simplest expression is for $j=\ha$, for which $B_{1/2} (x) = \tanh x$ and the potential becomes
\be
j={\ha}:\quad V(x) = (c_0 + c) x^2 - {2T} \Big( (1+x)\ln(1+x) + (1-x)\ln(1-x)\Big)\ .
\ee
In what follows we will use general properties of $V(x)$ for arbitrary $j$, reverting to $j=\ha$ for specific examples.

\begin{figure}[ht]
\centering
\usetikzlibrary{arrows.meta}

\begin{minipage}{0.49\textwidth}
\centering

\def\Temp{1.5}

\begin{tikzpicture}
\begin{axis}[
    width=7cm,
    height=6cm,
    axis lines=middle,
    axis line style={-{Latex[length=4mm,width=2.5mm]},thick},
    xmin=-1.05, xmax=1.25,
    ymin=-2.6, ymax=0.6,
    domain=-0.999:0.999,
    samples=400,
    xlabel={$x$},
    ylabel={$V(x)$},
    xlabel style={at={(axis description cs:.94,.87)},anchor=west},
    ylabel style={at={(axis description cs:0.35,.82)},anchor=south},
    xtick={-1,0,1},
    xticklabels={$-1$,$0$,$1$},
    ytick=\empty,
    enlargelimits=false,
    clip=false,
]

\addplot[
    blue,
    thick,
]
{2*x^2
 -2*\Temp*((1-x)*ln(1-x) + (1+x)*ln(1+x))};

\addplot[dashed,black,thick] coordinates {(-1,-2.5) (-1,0.08)};
\addplot[dashed,black,thick] coordinates {(1,-2.5) (1,0.08)};

\addplot[
    only marks,
    mark=*,
    mark size=2.5pt,
    blue
] coordinates {(0,0)};

\node[black,above right] at (axis cs:0,0) {$0$};

\end{axis}
\end{tikzpicture}

\end{minipage}
\hfill
\begin{minipage}{0.5\textwidth}
\centering

\def\Temp{0.68} 

\begin{tikzpicture}
\begin{axis}[
    width=7cm,
    height=6cm,
    axis lines=middle,
    axis line style={-{Latex[length=4mm,width=2.5mm]},thick},
    xmin=-1.05, xmax=1.25,
    ymin=-0.25, ymax=0.55,
    domain=-0.999:0.999,
    samples=400,
    xlabel={$x$},
    ylabel={$V(x)$},
    xlabel style={at={(axis description cs:.94,.37)},anchor=west},
    ylabel style={at={(axis description cs:0.35,.82)},anchor=south},
    xtick={-1,0 ,1},
    xticklabels={$-1$,$0$,$1$},
    ytick=\empty,
    enlargelimits=false,
    clip=false,
]

\addplot[
    blue,
    thick,
]
{2*x^2
 -2*\Temp*((1-x)*ln(1-x) + (1+x)*ln(1+x))};

\addplot[dashed,black,thick] coordinates {(-1,0) (-1,.435)};
\addplot[dashed,black,thick] coordinates {(1,0) (1,.435)};
\addplot[dashed,black,thick] coordinates {(0.854,0) (.854,.29)};
\addplot[dashed,black,thick] coordinates {(-0.854,0) (-.854,.29)};

\addplot[
    only marks,
    mark=*,
    mark size=2pt,
    black
] coordinates {(.854,0)};
\addplot[
    only marks,
    mark=*,
    mark size=2pt,
    black
] coordinates {(-.854,0)};

\node[below, black] at (axis cs:-0.8, 0.11) {$-x_{\max}$};
\node[below, black] at (axis cs:0.8, 0.11) {$x_{\max}$};

\addplot[
    only marks,
    mark=*,
    mark size=2.5pt,
    red
] coordinates {(0,0)};

\addplot[
    only marks,
    mark=*,
    mark size=2.5pt,
    blue
] coordinates {
     (-0.854,0.29)
     (0.854,0.29)
};

\node[black,above right] at (axis cs:0,0) {$0$};

\end{axis}
\end{tikzpicture}

\end{minipage}
\vskip -.1 cm
\captionsetup{
    labelfont={small},
    textfont={small}
}
\caption{The effective potential $V(x)$:  For $T>T_{c}$ (left) and  for $T<T_{c}$ (right).}
\label{effpotential}

\end{figure}
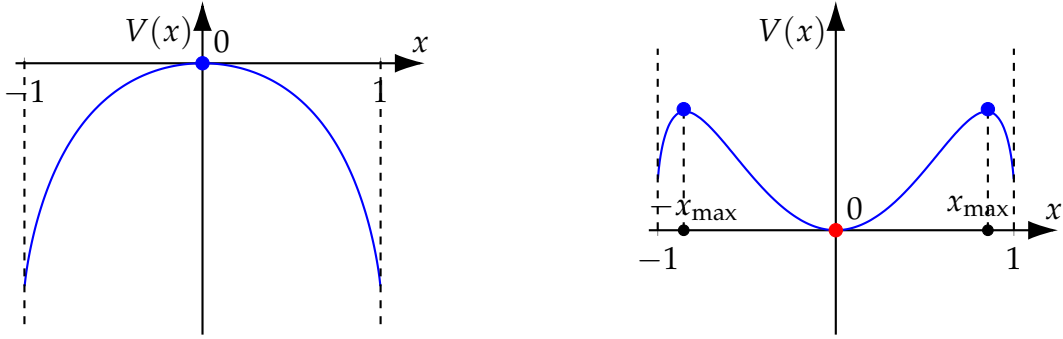

\no
For general $j$, the potential $V(x)$ near $x=0$ behaves as
\be
V(x) =
  {3(T_c -T) \over 2j(j+1)} x^2 - {a\ov 4}  T x^4 +\dots \ ,
\label{Vx0}
\ee
with the coefficient $a$ given by \eqn{ggga} and the critical temperature $T_c$ being
\be
\label{tcdef}
T_c = {2\over 3} j(j+1) (c_0 + c)\  .
\ee
Note that $T_c $ is the limit of $T_k$ in \eqn{Tk} as $m/k \to \infty$.
For $c_0 +c >0$, $T_c$ is positive. For $T>T_c$, the potential has a single maximum at $x=0$. For $T<T_c$,
it becomes an inverted "mexican hat" potential, with a local
minimum at $x=0$ and two symmetric maxima at $x = \pm x_\text{max}$. From \eqn{diffm1}, $x_\text{max}$ is the positive solution of
\be
x_\text{max} = B_j \big(2j^2 (c_0 + c) x_\text{max} /T\big)\ .
\label{xm}
\ee
For all temperatures, the potential near  $x=\pm 1$ behaves as 
\be
V(x)= c_0+c - {T\over j^2} \ln(2j+1) - {T\ov j} (1\mp x) \ln \big(1\mp x) + \dots  \ ,
\label{V1}\ee
so that $V'(\pm 1) = \mp \infty$.
In the purely paramagnetic case $c_0+c<0$, the potential has a single maximum at $x=0$ for all temperatures.
We will assume $c_0 +c >0$ from now on. The various cases are depicted in fig. \ref{effpotential}. 

\no
As we stressed in the beginning of this section, the continuum approximation is valid for configurations that change smoothly in $s$,
so that they can be considered as continuous functions of $s$ with derivatives $\ll 1$.
For motions of the fiducial particle away from $x=0$ and $x=\pm x_\text{max}$ this
is not strictly true, as \eqn{anharmonic} shows that the velocity of the particle is $\sqrt{2\big(E-V(x)\big)/c} \sim 1$. Nevertheless, the continuum
approximation remains qualitatively valid and will be useful in finding general properties of the magnet.

\no
The boundary conditions for the chain at $s=0$ and $s=m$ are deduced from \eqn{xsdis}, which requires $x_0 = x_{m+1} =0$.
In the continuum limit, they become the Dirichlet boundary conditions
\be
\label{diri}
x(0) = x(m) = 0 \ .
\ee

\no
The nature of the solutions of \eqn{xmotion} now becomes clear. To satisfy the boundary conditions  \eqn{diri}, the fiducial particle has to start at $x=0$
at $s=0$ and return to $x=0$ at $s=m$. Therefore, its motion must stay within $(-x_\text{max},x_\text{max})$, otherwise it could never return
to $x=0$. (The potential remains finite at $x=\pm1$, so the fate of the particle beyond $x_\text{max}$ is a bit unclear. Appendix \ref{stop} proves that,
indeed, $x$ must be in the range $|x|<x_\text{max}$.)
For $T>T_c$, the potential has a single global maximum at $x=0$ and the only solution is the singlet $x=0$, the particle sitting at the maximum.
For $T<T_c$ and $E<V(x_\text{max})$, the particle performs an anharmonic oscillation between the inflection points $\pm x_E$,
with $V(x_E) = E$, $0<x_E <x_\text{max}$. Values of $E$ larger than $V(x_\text{max})$ correspond to runaway solutions and are not relevant to the system.

\no
Small oscillations near $x=0$ become harmonic, and from \eqn{Vx0} their period is
\be
S_0 = 2\pi \sqrt{j(j+1) c\over 3(T_c -T)}\ .
\label{S0}
\ee
The period $S$ of anharmonic large oscillations is always greater than $S_0$, due to the negative fourth derivative of $V(x)$ in \eqn{Vx0},
and diverges as $E$ approaches the maximum $V(x_\text{max})$.
To obey the boundary conditions \eqn{diri}, the half-period $S/2$ must be $m/k$, $k=1,2,\dots$. Hence, for $S_0 >2m$
the boundary conditions cannot be satisfied for any $k$, and the only solution is the singlet. For values of $S_0 = 2m/k$, the small 
oscillations will reproduce a solution of the form
\be
x_k(s) = A \sin{\pi k s \over m}\ ,
\ee
while larger, anharmonic oscillation modes will give solutions with fewer than $k$ half-periods, all the way down to a single half-period. The temperatures
$T_k$ at which the mode $k$ appears are found by setting $S_0=2m/k$, giving, after using \eqn{S0}, 
\be
T_k = {2\over 3} j(j+1) \left(c_0 + c - {\pi^2 k^2 c \over 2m^2}\right) = T_c - {j(j+1) \pi^2 k^2 c \over 3 m^2}\ .
\label{Tks}
\ee
We see that $x_k(s)$ and $T_k$ agree with the exact modes \eqn{xsm} and \eqn{Tk} found in the previous section in the limit $m \gg 1$,
as long as $k\ll m$. The agreement is to ${\cal O}(k^2/m^2)$ which is related to the fact that in the continuum approximation we have kept 
up to two derivatives in the expansion in \eqn{derexpa}.

\begin{figure}[ht]
    \centering
    
    \begin{minipage}{0.49\textwidth}
        \centering
        \def\Temp{0.68}
        \begin{tikzpicture}
            \begin{axis}[
                width=7cm,
                height=6cm,
                axis lines=middle,
                axis line style={-{Latex[length=4mm,width=2.5mm]},thick},
                xmin=-1.05, xmax=1.25,
                ymin=-0.13, ymax=0.45,
                domain=-0.999:0.999,
                samples=400,
                xlabel={$x$},
                ylabel={$V(x)$},
                xlabel style={at={(axis description cs:.94,.37)},anchor=west},
                ylabel style={at={(axis description cs:0.35,.82)},anchor=south},
                xtick={-1,0,1},
                xticklabels={$-1$,$0$,$1$},
                ytick=\empty,
                enlargelimits=false,
                clip=false,
            ]
            
            \addplot[blue, thick] {2*x^2 - 2*\Temp*((1-x)*ln(1-x) + (1+x)*ln(1+x))};
            
            \addplot[dashed,gray, thick] coordinates {(-1,0) (-1,.435)};
            \addplot[dashed,gray,thick] coordinates {(1,0) (1,.435)};
            \addplot[dashed,gray,thick] coordinates {(0.854,0) (.854,.29)};
            \addplot[dashed,gray,thick] coordinates {(-0.854,0) (-.854,.29)};
            
            \addplot[only marks, mark=*, mark size=2pt, black] coordinates {(.854,0) (-.854,0)};
            \addplot[only marks, mark=*, mark size=2.5pt, red] coordinates {(0,0)};
            \addplot[only marks, mark=*, mark size=2.5pt, blue] coordinates {(-0.854,0.29) (0.854,0.29)};
            \node[black,above right] at (axis cs:0,0) {$0$};
            \node[below, black] at (axis cs:-0.8, 0.08) {$-x_{\max}$};
            \node[below, black] at (axis cs:0.8, 0.08) {$x_{\max}$};
            
            \draw[thick, red, -stealth] (axis cs: 0, 0.02) -- (axis cs: 0.25, 0.02);
            \draw[thick, red] (axis cs: 0.25, 0.02) -- (axis cs: 0.45, 0.02) 
                to[out=0, in=90] (axis cs: 0.55, 0) 
                to[out=270, in=0] (axis cs: 0.45, -0.02);
            \draw[thick, red, stealth-] (axis cs: 0.25, -0.02) -- (axis cs: 0.45, -0.02);
            \draw[thick, red] (axis cs: 0.25, -0.02) -- (axis cs: 0.05, -0.02) 
                to[out=180, in=270] (axis cs: 0, 0.02);
            
            \draw[thick, dashed,thick] (axis cs: -0.55, 0.17) -- (axis cs: 0.55, 0.17) node[right] {$E$};
            \draw[thick, dashed,thick] (axis cs: 0.55, 0.17) -- (axis cs: 0.55, 0);
            \draw[thick, dashed,thick] (axis cs: -0.55, 0.17) -- (axis cs: -0.55, 0);
            
            \end{axis}
        \end{tikzpicture}
    \end{minipage}
    \hfill
    \begin{minipage}{0.49\textwidth}
        \centering
        \begin{tikzpicture}
            \begin{axis}[
                width=7cm, height=6cm, 
                axis lines=middle,
                axis line style={-{Latex[length=3mm]},thick},
                xmin=0, xmax=5,
                ymin=0, ymax=1.3,
                xlabel={$s$}, ylabel={$x_1(s)$},
                xtick={0, 4}, xticklabels={$0$, $m$},
                ytick={1}, yticklabels={$x_{\max}$},
                enlargelimits=false,
            ]

            \node[black,above right] at (axis cs:0,0) {$0$};
\addplot[domain=0:4, samples=100, thick, red] {0.64 * (1 - ((x-2)^2/4)^1.2)};
\addplot[domain=0:4, samples=200, thick, dashed, red] {0.64 * sin(deg(x*pi/4))};
                \addplot[dashed, thick] coordinates {(0,1) (4,1)};
                   \addplot[dashed, thick] coordinates {(0,.64) (4,.64)};
            \end{axis}
        \end{tikzpicture}
    \end{minipage}

    \caption{\small{Left: Schematic path of the fiducial particle in the effective potential $V(x)$ for $T < T_{c}$ for the deformed $x_1$ mode,
    showing the energy level $E$ and its corresponding classical turning points. Right: The form of $x_1(s)$ for half a period, the red dashed line being the
    undeformed one corresponding to harmonic motion.}}
    \label{turnpotential}

\end{figure}
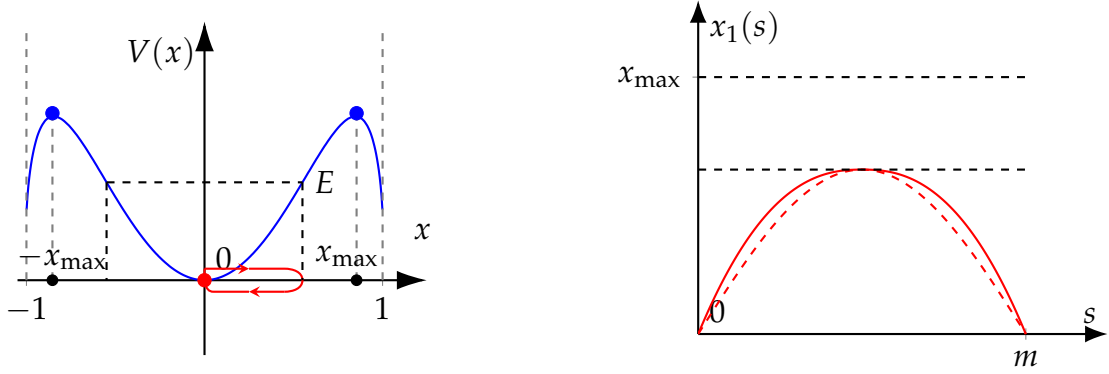

\no
At $T=T_1 \lesssim T_c$, the mode $x_1 (s)$ appears as a small oscillation of \eqn{anharmonic} with $E\simeq 0$. As $T$ decreases below $T_c$,
the potential $V(x)$ becomes steeper and $E$ must increase to maintain the same period of oscillation $2m$. In the process, the $k=1$ mode starts
being deformed, its slope being highest near $x(0)=0$, decreasing near the inflection point $x(m/2)$, and increasing again as it reaches $x(m)=0$. The 
motion, for half the period, is schematically depicted in fig. \ref{turnpotential}.


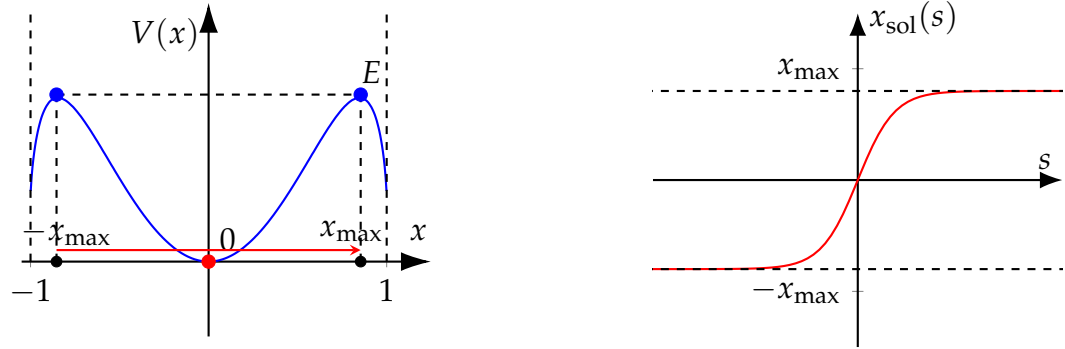
\begin{figure}[ht]
    \centering
    
    \begin{minipage}{0.49\textwidth}
        \centering
        \def\Temp{0.68}
        \begin{tikzpicture}
            \begin{axis}[
                width=7cm,
                height=6cm,
                axis lines=middle,
                axis line style={-{Latex[length=4mm,width=2.5mm]},thick},
                xmin=-1.05, xmax=1.25,
                ymin=-0.13, ymax=0.45,
                domain=-0.999:0.999,
                samples=400,
                xlabel={$x$},
                ylabel={$V(x)$},
                xlabel style={at={(axis description cs:.92,.31)},anchor=west},
                ylabel style={at={(axis description cs:0.35,.82)},anchor=south},
                xtick={-1,0,1},
                xticklabels={$-1$,$0$,$1$},
                ytick=\empty,
                enlargelimits=false,
                clip=false,
            ]
            
            \addplot[blue, thick] {2*x^2 - 2*\Temp*((1-x)*ln(1-x) + (1+x)*ln(1+x))};
            
            \addplot[dashed,black,thick] coordinates {(-1,0) (-1,.435)};
            \addplot[dashed,black,thick] coordinates {(1,0) (1,.435)};
            \addplot[dashed,black,thick] coordinates {(0.854,0) (.854,.29)};
            \addplot[dashed,black,thick] coordinates {(-0.854,0) (-.854,.29)};
            
            \addplot[only marks, mark=*, mark size=2pt, black] coordinates {(.854,0) (-.854,0)};
            \addplot[only marks, mark=*, mark size=2.5pt, red] coordinates {(0,0)};
            \addplot[only marks, mark=*, mark size=2.5pt, blue] coordinates {(-0.854,0.29) (0.854,0.29)};
            \node[black,above right] at (axis cs:0,0) {$0$};
            \node[below, black] at (axis cs:-0.8, 0.09) {$-x_{\max}$};
            \node[below, black] at (axis cs:0.8, 0.09) {$x_{\max}$};

            \draw[thick, red, -stealth] (axis cs: -0.854, 0.02) -- (axis cs: 0.854, 0.02);
            
            \draw[thick, dashed] (axis cs: -0.854, 0.29) -- (axis cs: 0.854, 0.29) node[above right] {$\hskip -.15 cm E$};
            
            \end{axis}
        \end{tikzpicture}
    \end{minipage}
    \hfill
    \begin{minipage}{0.49\textwidth}
        \centering
        \begin{tikzpicture}
            \begin{axis}[
                width=7cm, height=6cm,
                axis lines=middle,
                axis line style={-{Latex[length=3mm]},thick},
                xmin=-2.8, xmax=2.8,
                ymin=-1.2, ymax=1.2,
                xlabel={$s$}, 
                ylabel={$x_{\rm sol}(s)$},
                ylabel style={yshift=0.3cm},
                xtick={-4, 0, 4}, xticklabels={$-m$, $0$, $m$},
                ytick={-.8, .8}, yticklabels={$-x_{\max}$, $x_{\max}$},
                enlargelimits=false,
            ]
            
            \addplot[domain=-4:4, samples=100, thick, red] {0.64 * tanh(2*x)};
            
            \addplot[dashed, black,thick] coordinates {(-4,0.64) (4,0.64)};
            \addplot[dashed, black,thick] coordinates {(-4,-0.64) (4,-0.64)};
            
            \end{axis}
        \end{tikzpicture}
    \end{minipage}

    \caption{\small{Left: Path of the fiducial particle moving in the effective potential $V(x)$ for $T < T_{c}$ for energy $E=V_{\max}$ between
 the corresponding classical turning points $\pm x_{\rm max}$.\\ Right: The corresponding spatial trajectory $x(s)$ exhibiting a soliton (kink) profile.
 }}
    \label{turnpotential2}

\end{figure}

\no
As $T_c - T$ increases further, $E$ approaches the limiting value $V(x_\text{max})$ and the oscillation spends most of its "time" $s$
near the inflection point, deviating from it near the endpoints $s=0$ and $s=m$. Thus, $x_1(s)$ becomes a flat function with $x(s) \simeq x_\text{max}$ for
most of the domain of $s$, deviating from $x_\text{max}$ within a finite range near $x=0$ and $x=m$. The form of this mode can be expressed in terms of
the "kink soliton" solution of the particle with energy $E=V(x_\text{max})$ moving from $x=x_\text{max}$ to $x=-x_\text{max}$, depicted in fig. \ref{turnpotential2}.

\begin{figure}[htbp]
\centering

\begin{subfigure}[b]{0.32\textwidth}
\centering
\begin{tikzpicture}
\begin{axis}[
    axis lines = middle,
    xlabel = {$s$},
    ylabel = {$x_1(s)$},
    xmin = -0.5, xmax = 10.5,
    ymin = -1.25, ymax = 1.25,
    domain = 0:10,
    samples = 300,
    smooth,
    width = \textwidth, 
    height = 5.5cm,
    every axis x label/.style={at={(current axis.right of origin)},anchor=west},
    every axis y label/.style={at={(current axis.above origin)},anchor=south},
    xtick={0, 10},
    xticklabels={$0$, $m$},
    ytick={-1, 0, 1},
    yticklabels={$-x_{\max}$, $0$, $x_{\max}$},
    ymajorgrids = true,
    xmajorgrids = false,
    grid style = {dashed, black, thick}
]
    \addplot[blue, very thick] {tanh(x/0.45) * tanh((10-x)/0.45)};
\end{axis}
\end{tikzpicture}
\end{subfigure}\hfill
\hskip -.2 cm
\begin{subfigure}[b]{0.32\textwidth}
\centering
\begin{tikzpicture}
\begin{axis}[
    axis lines = middle,
    xlabel = {$s$},
    ylabel = {$x_2(s)$},
    xmin = -0.5, xmax = 10.5,
    ymin = -1.25, ymax = 1.25,
    domain = 0:10,
    samples = 300,
    smooth,
    width = \textwidth,
    height = 5.5cm,
    every axis x label/.style={at={(current axis.right of origin)},anchor=west},
    every axis y label/.style={at={(current axis.above origin)},anchor=south},
    yticklabels=\empty,
    xtick={0, 10},
 xticklabels={$0$, {$m\phantom{\hspace{20pt}}$}},
    ytick={-1, 0, 1},
    ymajorgrids = true,
    xmajorgrids = false,
    grid style = {dashed, black, thick}
]
    \addplot[blue, very thick] {tanh(x/0.45) * tanh((10-x)/0.45) * (-tanh((x-5)/0.45))};
\end{axis}
\end{tikzpicture}
\end{subfigure}\hfill
\hskip -.2 cm
\begin{subfigure}[b]{0.32\textwidth}
\centering
\begin{tikzpicture}
\begin{axis}[
    axis lines = middle,
    xlabel = {$s$},
    ylabel = {$x_3(s)$},
    xmin = -0.5, xmax = 10.5,
    ymin = -1.25, ymax = 1.25,
    domain = 0:10,
    samples = 400,
    smooth,
    width = \textwidth,
    height = 5.5cm,
    every axis x label/.style={at={(current axis.right of origin)},anchor= west},
    every axis y label/.style={at={(current axis.above origin)},anchor=south},
    xtick={0, 10},
    xticklabels={$0$, $m$},
    ytick={-1, 0, 1},
    yticklabels=\empty,
    ymajorgrids = true,
    xmajorgrids = false,
    grid style = {dashed, black, thick}
]
    \addplot[blue, very thick] {tanh(x/0.45) * tanh((10-x)/0.45) * tanh(sin(54*x)/0.4)};
\end{axis}
\end{tikzpicture}
\end{subfigure}
\caption{\small{Magnetization profiles for the $k=1, k=2,$ and $k=3$ modes along the chain length $m$, showing flat plateaus at $\pm x_{\max}$.}}
\label{magnprof}
\end{figure}
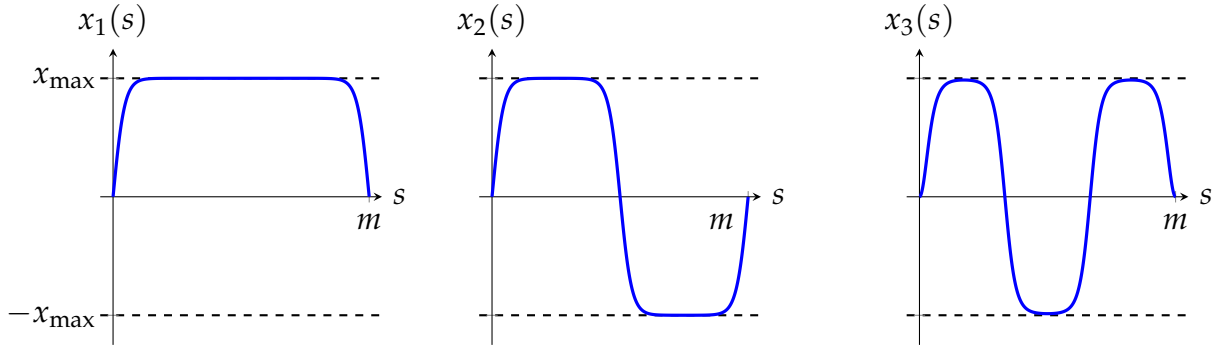

\no
The soliton mode satisfies
\be
x_\text{sol}' (s) = \sqrt{2\ov c\big(V(x_\text{max}) - V(x)\big)}\, , \quad x_\text{sol} (s) = -x_\text{sol} (-s)\, ,\quad x_\text{sol}(\pm \infty)= \pm x_\text{max}\ ,
\ee
the latter two conditions being a choice of time origin and direction of motion. The shape of $x_\text{sol} (s)$ is a smooth step around $s=0$ from
a plateau at $x=-x_\text{max}$ on the left to a plateau at $x=x_\text{max}$ on the right. The kink solution itself is not a physical state, since it does
not satisfy the vanishing boundary conditions at the end points $s=0,m$. Nevertheless, for $ E \lesssim V(x_{\rm max})$ the mode $x_1 (s)$ can be 
approximately expressed as
\be
x_1 (s) = x_\text{sol} \left({m\over 2} - \left| s- {m\over 2} \right| \right) \ ,\quad 0<s<m\ .
\ee
The range $S_t$ of the transition regions from $x=0$ to $x=x_\text{max}$ is estimated as
\be
S_t \sim S_0 \sim \sqrt{c \over T_c -T} \ . 
\ee
The transition is on a scale $\ll m$, and thus the continuum solution does not accurately reproduce the discrete transition values $x_s$,
but it captures the correct qualitative behavior.
The picture that emerges is a constant magnetization $x = x_\text{max}$ along the length of the chain, falling to zero near the edges $s=0,m$.
Higher modes $x_k (s)$, as $T$ dips below $T_k$, have a similar behavior: they consist of $k$ anharmonic half-oscillations, alternating from positive to
negative, with flat values $x=\pm x_\text{max}$ and transition regions from $x=0$ to $x=\pm x_\text{max}$ near $s=0,m$, and from $x=\pm x_\text{max}$ to
$x=\mp x_\text{max}$ near $s=p m/k$, $p=1,\dots,k-1$. The cases with $k=1,2,3$ and temperatures well below $T_k$ are depicted in fig. \ref{magnprof}.

%


\no
The states $x_k (s)$ become, essentially, a train of $k-1$ alternating kink solitons and antisolitons, with two additional half-(anti)solitons stuck to
the boundaries for a total soliton number zero, their mutual repulsion keeping them at equal distances.

\no
We proceed to study the stability of the above modes. The stability matrix $K=\b \bf{\Lambda C}$ now becomes a differential operator,
and its eigenvalue equation for $x=x_k(s)$ reads
\be
{2j^2\over T} B_j' \Big({2j^2} \big[(c_0 + c) x_k(s) + {c\over 2} x_k'' (s) \big]/T\Big)\big[(c_0+c) u(s)+ {c\over 2}u''(s)\big]= \l u(s)\ ,
\label{stabex}
\ee
with $u(s)$ the eigenstate.
The argument of $B_j' (\,\cdot\,)$ above is $2j^2(c_0+c) x_\text{max}/T$ for a range of values $s$ of size $\sim m$ and differs from it on a range of order $\ll m$, for all
$k \ll m$. Therefore, the operator $\b \bf{\Lambda C}$ is perturbatively close to
\be
{2j^2\over T} B_j' \Big({2j^2} (c_0 + c) x_\text{max}/T\Big)\big((c_0+c) u(s)+ {c\over 2}u''(s)\big)= \l u(s)
\label{stabxm}\ee
and has eigenvalues
\ba
\l_r = {2j^2\over T} B_j' \Big(2j^2 (c_0 +c) x_\text{max}/T \Big) \Big( (c_0+c) - {\pi^2 r^2 \over 2m^2} c \Big)
\, ,\quad r=1,2,\dots \ .
\ea
Stability requires $\l_r <1$. Since the last term is negative for all $r$, we see that the state $x_k (s)$ will be stable if
\be
{2j^2(c_0+c)\over T} B_j' \Big(2j^2 (c_0 +c) x_\text{max}/T \Big) <1\ .
\ee
Differentiating \eqn{xm} with respect to $T$ (taking into account that $x_\text{max}$ depends on $T$), 
\be
{2j^2(c_0+c)\over T}  B_j' \Big({2j^2(c_0+c)x_\text{max}/ T}\Big) = 1 +  {x_\text{max} \over T (\partial x_\text{max} / \partial_T)- x_\text{max}}\ .
\label{stamax}\ee
Since $x_\text{max}$ is a decreasing function of $T$,
the second term is negative and the stability condition is identically satisfied.
Therefore, states $x_k(s)$ are stable for all values of $T<T_k$, for which $x_k(s)$ has approached its asymptotic "soliton train" form. This is to be contrasted with the result for states perturbatively close to the singlet,
examined before, which became unstable as $T$ dipped below $T_1$. 
\no
The above result will receive $1/m$ corrections, but will remain valid for modes $k\ll m$ and temperatures low enough for these modes to flatten to
$x= \pm x_\text{max}$. Combining with the results
of the previous section on the discrete system for temperatures close to $T_c$, we conclude that, indeed, modes $x_k (s)$ become stable below
some critical temperature $T_{c,k} <T_k$.

\no
Finally, we can estimate the number of stable modes $k$ at temperature $T<T_c$. The maximum $k$ corresponds to
excitations near the bottom of the potential, which from \eqn{S0} and $S_0 = m/k$ correspond to $k = k_\text{max}$,
\be
k_\text{max} = {m\over \pi} \sqrt{3(T_c -T)\over 2j(j+1)c}\ .
\ee
This mode, being perturbatively close to $x=0$, will be unstable. As $k$ decreases from $k_\text{max}$, the modes $x_k (s)$
increase in size and approach the saturation value $x_\text{max}$, for which stability holds. Therefore, the
number of stable modes $K$ at temperature $T$ is of order
\be
K \lesssim m \sqrt{T_c -T \over j(j+1) c}\ .
\label{kmax}\ee
As a check, for $c=0$ the system becomes a set of $m$ uncoupled magnets, each being unmagnetized if $T>T_c$,
with $T_c = 2j(j+1) c_0/3$ the usual Curie temperature, and having magnetization $\pm x_\text{max}$ for $T<T_c$.
Choosing their signs arbitrarily we have a number of states $2^m \sim \infty$, in accordance with \eqn{kmax}.

\no
We note that for temperatures $T$ close to $T_c$, $x(s)$ will remain small and the potential can be expanded around $x=0$ taking the form 
\eqn{Vx0} with $T$ in the second term replaced by $T_c$, i.e. 
\be
\label{hh3}
V(x) = {3(T_c - T)\over 2j(j+1)}\, x^2 - {9(2j^2+2j+1) T_c \over 40 j(j+1)^3}\, x^4 + \dots\ ,
\ee
which has the form of a (flattened) mexican hat for $T<T_c$. The soliton solution $x_\text{sol} (s)$, obtained by solving  \eqn{anharmonic} for this
approximate potential and $E=V(x_{\rm max}) $  can now be found explicitly, as
\be
\label{xsol}
x_\text{sol} (s) = x_\text{max} \tanh\hb(q\, s),\quad 
x_\text{max} = (j+1)\sqrt{10 (T_c -T) \over 3(2j^2+2j+1)T_c} ~,\quad q = \sqrt{3(T_c -T) \over 2 j(j+1) c} 
\ee
and for $j=\ha$ it simplifies to
\be
x_\text{sol} (s) = \sqrt{3 {T_c -T \over  T_c}} \tanh \bigg(\sqrt{2{ T_c-T \over c}} s\bigg) \ .
\ee

\subsection{Modified boundary conditions and boundary states}
\label{modbc}

\no
Modified boundary conditions, different  than Dirichlet, can arise by introducing defects in the magnetic chain by modifying the couplings near the edges
$s=1$ and $s=m$ in the discrete chain. Focusing on $s=1$, consider the simplest case in which the coupling between $x_1$ and $x_2$ is modified to $c'$.
(Note that the sign of $c'$ can be flipped by redefining $x_1 \to -x_1$.) 

\no
This modification will lead to modified boundary conditions in the continuum limit in the
regime where $x'(s) \ll 1$.
To determine them, consider the difference equation in the linearized limit \eqn{xex}, valid for all $x_s$ except for $s=1,2$, since it then involves
the modified coupling constant $c'$. The equations for $s\geqslant 3$ remain valid and determine the solution for all $x_s$ with $s\geqslant 2$, which
is of the form \eqn{xlin}, 
\be
x_s = A \sin (s \th +\d)\ , \qquad \cos \th = 1- {3(T_c-T) \over 2j(j+1)c}\ ,
\ee
expressed in terms of the critical temperature $T_c$ \eqn{tcdef}.
The value of $x_1$, however, will be modified. The equations for $s=1$ and $s=2$ read, respectively,
\be
\label{c1c2}
c'x_2 = 2 c \cos\th \, x_1   \ ,\qq c' x_1 + c x_3 = 2  c \cos\th \, x_2   \ .
\ee
Eliminating $x_1$, we obtain
\be
{x_3\ov x_2} = {\sin (3 \th + \d)\ov \sin(2 \th +\d)} = {4 c^2 \cos^2\th - c'^2\ov 2 c^2\cos\th}\ ,
\ee
and using the trigonometric identity
\be
{\sin (3 \th + \d)\ov \sin(2 \th +\d)} = \cos\theta + \sin\theta \frac{\cot\delta - \tan 2\theta}{1+\tan 2\theta  \cot\delta} \ ,
\ee
we find for the phase shift
\be
\label{phaseshift}
\cot \d =  
{c^2  \csc 2 \th \ov c'^2-c^2}- \cot 2\th\ .
\ee
Extending the solution $\sin(\th s+\d)$ to $s=0$, and assuming $s$ is a continuous variable,
the boundary condition for $x(s)$ is 
\be
x'(0) = \mu\, x(0)\ ,\qq \mu =\th \cot\d  =  {c^2  \th\csc  2\th \ov c'^2-c^2}- \th\cot 2\th \ .
\label{bc}
\ee
This boundary condition is of the Robin type, involving both the field and its derivative. Particular cases are
\be
\begin{split}
&
{\rm  Dirichlet,}\ \ x(0)=0: \quad  c'=c \ ({\rm as\ expected})\ ,
\\
& {\rm   Neumann,}\ x'(0)=0: \quad\bb c'^2 = c^2 (1+ \sec 2\theta) \ . 
\end{split}
\ee
For $c'=0$, the site $x_1$ decouples from the rest of the chain, which remains the same as the one for $c'=c$ with $m \to m-1$.
The boundary condition becomes $x'(0) = - \theta \cot\theta\, x(0)$, but that implies, due to \eqn{phaseshift}, that the phase shift becomes $\d=-\th$
with the end  effect of making $x(s)$ equal to $x(s-1)$ of the chain with $c' =c$ and no phase shift, as expected.

\no
The regime $x'(s)\ll 1$ where the continuum limit becomes a good approximation corresponds to $\th \ll 1$ and temperatures close to $T_c$.
In this case, the boundary condition becomes
\be
x'(0) =\mu_0\,\hb  x(0)\ , \quad  \mu_0 = {2c^2 - c'^2\over 2(c'^2 -c^2)}\ .
\label{muco}
\ee
Away from it, \eqn{bc} becomes an approximation and gives $\mu$ a temperature dependence (through $\th$).
Nevertheless, the continuum approach will give us qualitative results on the behavior of the system even at temperatures away from $T_c$.

\no
Similar conditions arise at $s=m$, with an extra minus sign in $x'(0)$, upon setting the coupling between $x_{m-1}$ and $x_m$ to $c'$.
We also note that modifying the coupling constants between any small ($\ll \bb m$) number of spins near $s=0$ would lead to the same
mixed boundary conditions in the continuum limit, with the coefficient $\mu$ a function of the modified coupling constants.


\subsubsection{Modified boundary condition effects in the discrete linearized system}

The physical effects of the modified boundary conditions depend on the size and sign of $\mu$ and on the temperature. Returning to the discrete
system in the linearized approximation, we will focus on the case of
Robin boundary conditions \eqn{bc} at $s=0$ and standard Dirichlet ones at the other end of the chain ($x_{m+1}=0$), and will
consider separately the domains $T<T_c$ and $T>T_c$.

\begin{figure}[ht]
    \centering
    \begin{tikzpicture}
    \begin{axis}[
        legend pos=outer north east,
        legend style={draw=none, font=\small, fill=white, fill opacity=0.8, text opacity=1, row sep=1.5ex },
        width=12cm, height=8.5cm,
        trig format plots=rad, 
        xmin=0, xmax=3.28,
        ymin=-3.5, ymax=3.5, 
        ytick={0}, 
        xtick={0, 0.785398, 1.570796, 2.356194, 3.141593},
        xticklabels={{0}, {\hspace{0pt} $\frac{\pi}{4}$}, {\hspace{0pt} $\frac{\pi}{2}$}, {\hspace{0 pt}$\frac{3\pi}{4}$}, {$\pi$}},
        extra y ticks={1.5492, -1.5492},
        extra y tick labels={$c'_{\text{cr}}$, $-c'_{\text{cr}}$},
        extra y tick style={grid=major, major grid style={thick, black, dashed}},
        grid=none,
        axis lines=middle,
        axis line style={thick},
        xlabel={$\theta$},
        ylabel={$c'$},
        every axis x label/.style={at={(current axis.right of origin)}, anchor=west, xshift=.6 mm},
        every axis y label/.style={at={(current axis.above origin)}, anchor=south, yshift=.6 mm},
        every axis plot/.append style={line cap=round, line join=round}, 
    ]

        \def\myfunc{sqrt(abs(2 * cos(x) * sin(6*x) / sin(5*x)))}

        \addplot[thick, blue, samples=150, domain=0.001:0.523598, restrict y to domain=-5:5, forget plot] {\myfunc};
        
        \addplot[thick, blue, samples=150, domain=0.630:1.047197, restrict y to domain=-5:5, forget plot] {\myfunc};
        
        \addplot[thick, blue, samples=200, domain=1.259:1.882, restrict y to domain=-5:5, forget plot] {\myfunc};
        
        \addplot[thick, blue, samples=150, domain=2.094395:2.511, restrict y to domain=-5:5, forget plot] {\myfunc};
        
        \addplot[thick, blue, samples=150, domain=2.617993:3.141, restrict y to domain=-5:5, forget plot] {\myfunc};

        \addplot[thick, blue, samples=150, domain=0.001:0.523598, restrict y to domain=-5:5, forget plot] {-\myfunc};
        
        \addplot[thick, blue, samples=150, domain=0.630:1.047197, restrict y to domain=-5:5, forget plot] {-\myfunc};
        
        \addplot[thick, blue, samples=200, domain=1.259:1.882, restrict y to domain=-5:5, forget plot] {-\myfunc};
        
        \addplot[thick, blue, samples=150, domain=2.094395:2.511, restrict y to domain=-5:5, forget plot] {-\myfunc};
        
        \addplot[thick, blue, samples=150, domain=2.617993:3.141, restrict y to domain=-5:5, forget plot] {-\myfunc};

        
           \addlegendentry{$c' = 2.2c$}   
        \addplot[thick, green!50!black, dashed] {2.2};
                \addlegendentry{$c' = 0.8c$}       
                \addplot[thick, red, dashed] {0.8};
        
%

        \node[anchor=south east, font=\small, fill=white, fill opacity=0.8, text opacity=1] 
            at (rel axis cs: 1.01, 0.05) {$c'_{\text{cr}} = \sqrt{\frac{12}{5}}c$};

   \end{axis}
    \end{tikzpicture}
    \caption{\small{Plot of the left hand side of \eqn{ghhr} for $m=6$ and representative $c'$ values.}}
\label{ratio_m6}
\end{figure}
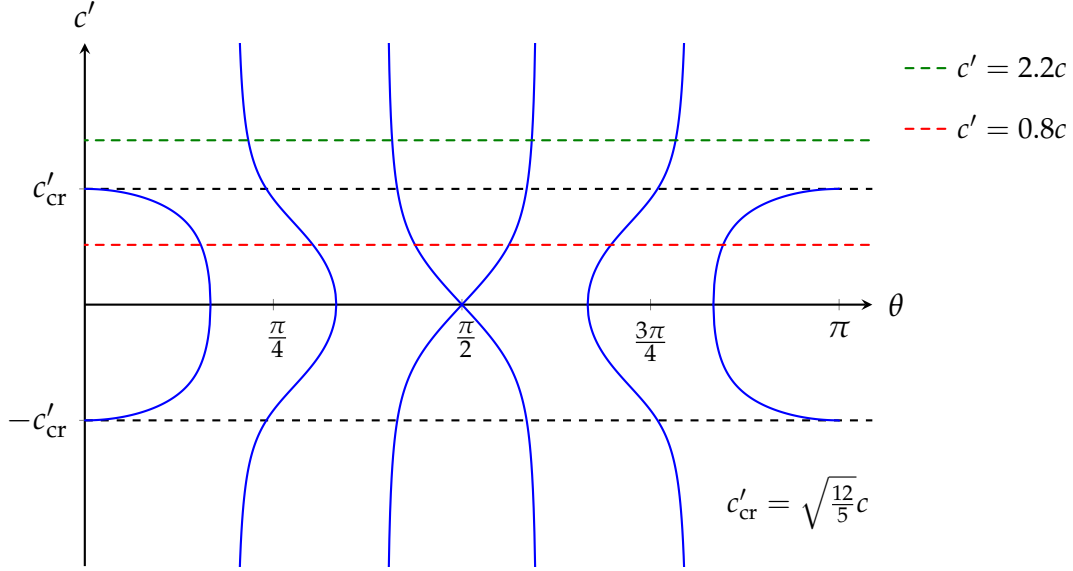

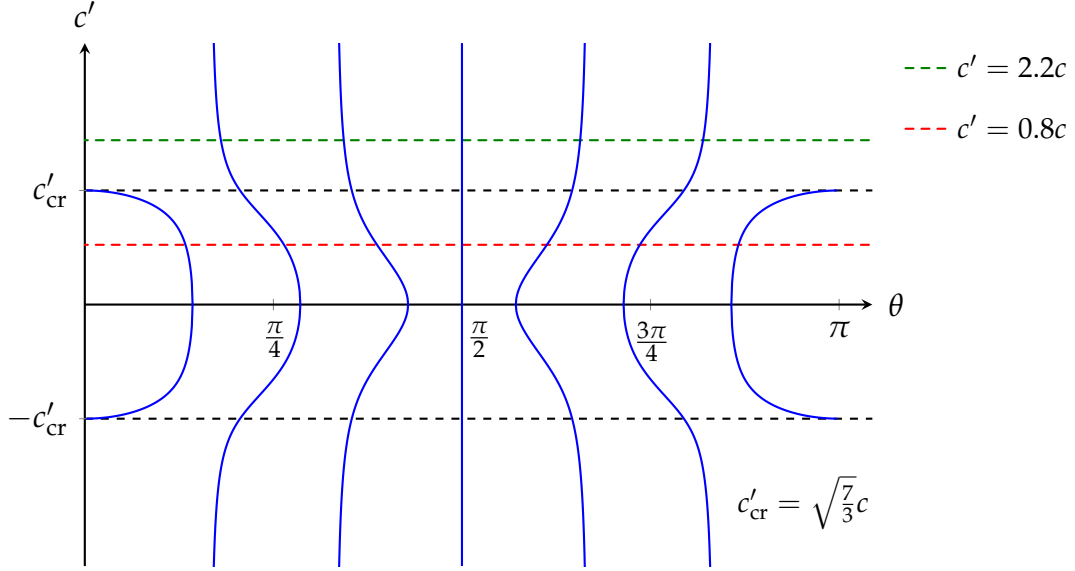
\begin{figure}[ht]
    \centering
    \begin{tikzpicture}  
    \begin{axis}[
        legend pos=outer north east,
        legend style={draw=none, font=\small, fill=white, fill opacity=0.8, text opacity=1, row sep=1.5ex },
        width=12cm, height=8.5cm,
        trig format plots=rad, 
        xmin=0, xmax=3.28,
        ymin=-3.5, ymax=3.5, 
        ytick={0}, 
        xtick={0, 0.785398, 1.570796, 2.356194, 3.141593},
        xticklabels={{0}, {\hspace{-2pt} $\frac{\pi}{4}$}, {\hspace{10pt} $\frac{\pi}{2}$}, {\hspace{1 pt}$\frac{3\pi}{4}$}, {$\pi$}},
        extra y ticks={1.5276, -1.5276},
        extra y tick labels={$c'_{\text{cr}}$, $-c'_{\text{cr}}$},
        extra y tick style={grid=major, major grid style={thick, black, dashed}},
        grid=none,
        axis lines=middle,
        axis line style={thick},
        xlabel={$\theta$},
        ylabel={$c'$},
        every axis x label/.style={at={(current axis.right of origin)}, anchor=west, xshift=.6 mm},
        every axis y label/.style={at={(current axis.above origin)}, anchor=south, yshift=.6 mm},
        every axis plot/.append style={line cap=round, line join=round}, 
    ]

        \def\myfunc{sqrt(abs(2 * cos(x) * sin(7*x) / sin(6*x)))}

         \addlegendentry{$c' = 2.2c$}   
        \addplot[thick, green!50!black, dashed] {2.2};
                \addlegendentry{$c' = 0.8c$}       
                \addplot[thick, red, dashed] {0.8};
       \addplot[thick, blue, samples=150, domain=0.01:0.448799, restrict y to domain=-5:5] {\myfunc};
        
        
        \addplot[thick, blue, samples=150, domain=0.524:0.897598, restrict y to domain=-5:5, forget plot] {\myfunc};
        
        \addplot[thick, blue, samples=150, domain=1.048:1.346397, restrict y to domain=-5:5, forget plot] {\myfunc};
        
        \addplot[thick, blue, samples=150, domain=1.795196:2.093, restrict y to domain=-5:5, forget plot] {\myfunc};
        
        \addplot[thick, blue, samples=150, domain=2.243995:2.617, restrict y to domain=-5:5, forget plot] {\myfunc};
        
        \addplot[thick, blue, samples=150, domain=2.692794:3.13, restrict y to domain=-5:5, forget plot] {\myfunc};

        \addplot[thick, blue, samples=150, domain=0.01:0.448799, restrict y to domain=-5:5, forget plot] {-\myfunc};
        
        \addplot[thick, blue, samples=150, domain=0.524:0.897598, restrict y to domain=-5:5, forget plot] {-\myfunc};
        
        \addplot[thick, blue, samples=150, domain=1.048:1.346397, restrict y to domain=-5:5, forget plot] {-\myfunc};
        
        \addplot[thick, blue, samples=150, domain=1.795196:2.093, restrict y to domain=-5:5, forget plot] {-\myfunc};
        
        \addplot[thick, blue, samples=150, domain=2.243995:2.617, restrict y to domain=-5:5, forget plot] {-\myfunc};
        
        \addplot[thick, blue, samples=150, domain=2.692794:3.13, restrict y to domain=-5:5, forget plot] {-\myfunc};



        \node[anchor=south east, font=\small, fill=white, fill opacity=0.8, text opacity=1] 
            at (rel axis cs: 1.01, 0.05) {$c'_{\text{cr}} = \sqrt{\frac{7}{3}}c$};

        \draw[thick, blue] (axis cs: 1.570796, -3.5) -- (axis cs: 1.570796, 3.5); 

   \end{axis}
    \end{tikzpicture}
    \caption{\small{Plot of the left hand side of \eqn{ghhr} for $m=7$ and representative $c'$ values. 
    The vertical line at $\theta = \frac{\pi}{2}$ represents a solution existing for all values of $c'$.}}
\label{ratio_m7}
\end{figure}

\no
For $T<T_c$ and for threshold modes with $x_k(s)\ll 1$, $\mu$ modifies the allowed values for $\th$ and $\d$ in \eqn{xsm}.
The condition $x_{m+1}=0$ implies $\d = -(m+1) \th$. 
Using \eqn{phaseshift} and a trigonometric identity, we obtain the transcendental equation for $\theta$
\be
\label{ghhr}
\pm  \sqrt{2\cos\theta \, \sin m\theta \over \sin (m-1)\theta} = {c'\ov c} \ .
\ee
This equation is valid when $\th$ is real, that is, for
\be
\label{trig}
\cos\th = {3 (T-T_c)\ov 2 j(j+1) c} +1\, ,\quad T_c-{4\ov 3} j(j+1) c < T < T_c\ .
\ee
The cases $m=6$ and $m=7$ are representative of the generic odd and even values of $m$ and are depicted in figs. \ref{ratio_m6}
and \ref{ratio_m7}, respectively.


\no
Each solution for $\th$ corresponds to a critical temperature below which the corresponding state appears, as in the undeformed case.
Since $\th\to \pi-\th$ is an invariance of \eqn{ghhr}, these solutions come in pairs $(\th,\pi-\th)$. The self-invariant value $\th=\pi/2$ is a solution for odd $m$ for all values of $c'\neq 0$.
The configuration is $x_s = A\sin(m+1-s)\pi/2$, so $x_2 = 0$ for odd $m$, satisfying the first of \eqn{c1c2}, while the second one yields $x_1 = -c x_3/c'$.
For even $m$, $x_2 = 0$ would imply $A=0$ eliminating the solution.
For $c'=c$ we recover the solutions $\th_k= {\pi k\ov m+1}$, $k=1,2,\dots m$ given in \eqn{xsm} and the corresponding temperatures are just the $T_k$'s given in \eqn{Tk}.
For larger $c'$, there is a critical value $c'_\text{cr}$ 
\be
c'_\text{cr} = \sqrt{2m\over m-1}\, c\ ,
\ee
above which the solutions near $\th=0$ and $\th=\pi$ disappear. Therefore, in the range $|c'|>c'_\text{cr}$
there are $m-2$ solutions. The two missing solutions are recovered by taking $\theta$ to be complex; specifically,
\be
\begin{split}
\th &= i\kappa \ , \qq ~~\cosh \kappa = {3(T-T_c) \over 2j(j+1)c} +1\, , \quad T>T_c \ ,\cr
\th &= \pi +i\kappa \ , \ \  \  \cosh \kappa = {3(T_c-T) \over 2j(j+1)c} -1\, , \quad  T<T_c - {4\over 3}j(j+1)c \ .
\end{split}
\label{hyper}
\ee
In either case, the equation for $\kappa$ becomes
\be
\label{ghhr2}
\pm  \sqrt{2\cosh\ka \, \sinh m\ka \over \sinh (m-1)\ka} = {c'\ov c} \ .
\ee
and has a unique positive solution for
\be
c'^2 > c_{\rm cr}^2 =  {2m \over m-1} c^2\ .
\label{cpcr}
\ee
The solutions for $x_s$ are $x_1=A (c / c')  \sinh m\k$, and, for $s=2,\dots , m$,
\be
\begin{split}
x_s &= \, A \sinh \k(m+1-s)\ , \qq\qq~~ \text{for}~~~ T>T_c \ ,\cr
 &= A(-1)^{s-1} \sinh \k(m+1-s)\ , \quad \text{for}~~~ T<T_c - {4\over 3}j(j+1)c \ .  
\label{xbound}\end{split}
\ee
These are boundary states, decaying exponentially away from $s=1$ with a rate $\k$.

\no
We conclude that for $|c'| > c'_\text{cr}$, 
the magnet in the linearized approximation will have a boundary state, a smooth one at a temperature $T_B$ and an alternating one at a temperature
$\tilde T_B$. The corresponding critical temperatures $T_B,\tilde T_B$ are determined by \eqn{hyper} and \eqn{ghhr2}. In the limit of a long chain
$m\gg 1$,  $T_B$ and $\tilde T_B$ can be found exactly.
Indeed, in that limit the solution of \eqn{ghhr2} is $e^\kappa = \sqrt{c'^2/c^2-1}$, and \eqn{hyper} gives
\be
\label{TBTL}
\begin{split}
&
T_B = T_c + {j(j+1)\over 3}\bigg({c'^2 \over \sqrt{c'^2 - c^2}}- 2c\bigg) > T_c\ ,
 \\
 &\tilde T_B = T_c - {j(j+1)\over 3}\bigg({c'^2 \over \sqrt{c'^2 - c^2}}+2 c\bigg) < T_c -{4\ov 3} j(j+1) c\ .
\end{split}
\ee
Note that $T_B$ lies above, and $\tilde T_B$ lies below, the critical temperatures $T_k$
at which extended states of the chain start to appear, i.e., $\tilde T_B<T_k<T_B$.
This  can be readily checked by noting, using  \eqn{Tk}, that $T_c-{4\ov 3} j(j+1) c < T_k < T_c$.

\no
The boundary state for $T<T_B$ with $\ka \ll1$ varies smoothly with $s$ and can be described in the continuum approximation. These conditions require $T\lesssim T_B$ and $0<-\mu_0 \ll 1$  (possible since, due to \eqn{cpcr},  ${1\ov m+1}\leqslant -\mu_0 \leqslant \ha $). In fact, its qualitative properties are captured by the continuum approximation even away from these values.
The boundary state for $T<\tilde T_B$, on the other hand, is outside of the range of validity of the continuum approximation, as it is oscillatory at a scale $\Delta s=1$ and does not admit the standard continuum form. Instead, it is captured by the dual continuum limit defined in subsubsection \ref{dualcont}.
In Appendix \ref{fullmonty} we prove the existence of the boundary state in the exact system, without the
linearized or continuum approximations.

\no
We conclude by checking a couple of special cases. For $c'=0$, as seen from \eqn{c1c2}, $x_1$ decouples and the system reduces to a chain with one less atom and usual
(Dirichlet) boundary conditions plus an isolated atom. The values of $\th$ in this case, from \eqn{ghhr}, are
\be
\th = {\pi k \over m} \ , \quad k=1,\dots,m-1 \ , \quad \th={\pi\over 2} \ .
\ee
The values for $k=1,\dots,m-1$ are those of an undeformed chain with $m-1$ atoms, while $\th = {\pi\over 2}$ corresponds to the critical
temperature of the isolated atom
\be
T_\text{isol} = {2\over 3} j(j+1) c_0 = T_c - {2 \over 3}j(j+1) c\ .
\ee
The value $\th = {\pi \over 2}$ is doubly degenerate for even $m$, corresponding to
independent excitations of $x_1$ and the rest of the chain.
For $c'=\infty$ and \eqn{c1c2}, both $x_1$ and $x_2$ are frozen to zero to ensure a finite free energy, and the system reduces to a
standard chain with two fewer atoms. The solutions, from \eqn{ghhr}, are
\be
\theta = {\pi k \over m-1} \ , \quad k=1,\dots,m-2 \ .
\ee
The two would-be boundary states vanish in this case. 

\no
We also note the expression for $\theta \simeq 0$ near the critical value $c' \lesssim c'_\text{cr}$ 
\be
 \label{eccc}
\th = \bigg({18 (m-1)\ov m(m+1)^2} \bigg)^{1/4} \sqrt{c'_{\rm cr}-c' \over c}+\dots \ . 
\ee
 The corresponding critical temperature is 
  \be
 T_{1,{\rm def}} = T_c - {j(j+1)\over m+1}\sqrt{2{m-1\over m}} (c_\text{cr} - c') + \dots \ .
 \ee
This is the deformed temperature $T_1$ for $c'=c$ in \eqn{Tk}  since, as seen from figs. \ref{ratio_m6} and \ref{ratio_m7}, only the first 
 solution  for $\th$ approaches zero at $c'=c'_{\rm cr}$, as in  \eqn{eccc}.

\subsubsection{Modified boundary condition effects in the continuum limit}

In the continuum limit the effect of Robin boundady conditions is that the fiducial particle does not start at $x=0$ but rather at another value
such that $x'(0) = \mu x(0)$, keeping the standard Dirichlet boundary condition $x(m)=0$ at the other end of the chain. 
The effects on the solution depend on the sign of $\mu$ and the temperature.
In the following analysis we will take $\mu$ to be independent of the temperature and the state of the system, as in \eqn{muco}.

\no
For $\mu>0$, the effects are qualitatively minimal. For $T>T_c$, any nonzero $x(0)$ would lead to a runaway solution, leaving the singlet
$x(s)=0$ as the only state. For $T<T_c$, the solutions would be of the form depicted in figs. \eqn{turnpotential} and \eqn{magnprof}, but starting at
a point between $x=0$ and its maximal value. The ratio $\mu = x' /x$ in this range can take all positive values, from $\mu=\infty$ at $x=0$ to
$\mu=0$ at maximum $x$. The effect is that the motion is truncated, and to reach $x=0$ at $s=m$ its period $S$ must be larger than the one for $\mu=\infty$,
and the value of $\theta=2\pi/S$ decreases. This is consistent with the discrete linearized analysis when $\mu$ is small and $m\gg1$, where both approaches are valid:  
from \eqn{muco}, $\mu\gtrsim 0$ means $c^2 < c'^2 < 2c^2$, so $c'$ is below its critical value. Also, in the continuum only modes with $\theta \ll 1$ are relevant.
As is clear from figs. \ref{ratio_m6} and \ref{ratio_m7}), for modes with $\theta <{\pi\over 2}$
the corresponding values of $\th$ are indeed smaller than the ones for $c'=c$.

\no
The situation is more interesting for $\mu <0$. For $T<T_c$, the situation is similar to the one for $\mu>0$, except now the motion is expanded rather than
truncated, and its period must be smaller than the one for $\mu =\infty$, so $\theta$ increases. Indeed, small negative $\mu$ means that $c' \gg c$, and
figs.~\eqn{ratio_m6} and \eqn{ratio_m7} show that for such values of $c'$ the values of $\th$ increase.

\no
For $T>T_c$, a new situation arises: other than the singlet $x(s)=0$, the fiducial particle can start at some $x(0)>0$ with a velocity $x'(0) = \mu\, x(0) <0$
and "creep" to the maximum of the potential at $x(m)=0$. This state is nonzero near the boundary at $s=0$ and decays to zero away from it:
it is a boundary state. The energy $E$ should be slightly larger than $0$ so that the particle
can reach the point $x=0$ at finite "time" $s=m$. For large $m$, we can take $E=0$, since the tail of the bound state will decay long before $s=m$.
For a motion with $E=0$, starting from $x(0)>0$ and moving towards $x=0$, we have
\be
x'(0) = -\sqrt{-{2\over c}V\big(x(0)\big)}\ .
\ee
The boundary condition becomes
\be
\mu = {x'(0) \over x(0)} = - {\sqrt{-{2\over c}V\big(x(0)\big)} \over x(0)} <0 ~ .
\quad  
\mu^2 = -{2V\big(x(0)\big)\over c x(0)^2}\ .
\ee
The starting point $x(0)$ must satisfy the above condition.
Since $-V(x)$ increases faster than quadratically, the right hand side has a minimum at $x(0) \to 0$ equal to
\be
\left. -{2V\big(x\big)\over c x^2}\right|_{x \to 0} = {3(T-T_c)\over j(j+1)c}
\ee
and a maximum as $x(0)$ reaches the value $1$, from \eqn{V1} equal to 
\be
-{2\over c}V(1)= {2T\over j^2 c} \ln(2j+1) -{3 T_c \over j(j+1) c}\ .
\ee
Therefore, the boundary condition can be satisfied for all temperatures for which
$\mu^2$ is between the maximal and minimal value, that is,
\be
{3(T-T_c)\over j(j+1)c}< \mu^2 < {2T\over j^2 c} \ln(2j+1) -{3 T_c \over j(j+1)c}\ ,
\ee
which implies
\be
T_b < T < T_B\ ,
\ee
with
\be
T_B = T_c + {j(j+1)c\mu^2 \over 3}~,\quad T_b = {3j\over 2(j+1) \ln(2j+1)} T_B\ .
\label{Bcon}\ee
Note that $T_B>T_c$. For $T_b$ to be relevant it must also be bigger than $T_c$. This requires
\be
\mu^2 > \mu_c^2 = \left({2\ln(2j+1)\over j^2} - {3\over j(j+1)}\right) {T_c \over c}
\label{muc}\ee
For $\mu^2<\mu_c^2$ there is no lower temperature $T_b$.
The picture that emerges is that, as $T$ increases above $T_c$, originally only the singlet state exists (if $-\mu>\mu_c$).
As $T$ crosses the critical temperature $T_b$,
a boundary state appears along with the singlet, and persists until $T=T_B$, beyond which the singlet becomes again the only state.

\no
The amplitude of the boundary state is zero at $T=T_B$, and very small just below it, so the exact expression for $T_B$ is captured by the linearized
discrete analysis \eqn{TBTL}.
To check the agreement of the linear discrete and continuum limits, we focus on the situation in which the coefficient $\mu$ is small, where $\kappa$ 
in \eqn{hyper} is small and both approaches are valid. Setting
\be
c'^2 = 2c^2 (1+\delta)~, \qq 0<\d\ll 1 \ ,
\ee
equation \eqn{muco} gives $\mu = -\d$, while \eqn{TBTL} gives
\be
T_B = T_c + {j(j+1)\over 3} c \d^2 = T_c + {j(j+1) c \mu^2 \over 3}\ ,
\label{smallm}\ee
in accordance with the continuum equation \eqn{Bcon}.

\no
The critical temperature $T_b$, on the other hand, cannot be captured by the linear analysis, as the boundary state disappears
nonperturbatively below $T_b$, and \eqn{muc} shows that we cannot take $\mu$ to be small.
Also, the specific realization of boundary conditions with a single coupling $c'$ differing from $c$ is one of many that would lead to the
same continuum limit. As shown in Appendix \ref{fullmonty}, there is no lower critical temperature $T_b$ in the exact discrete system with a single $c' \neq c$.
It remains an open question whether $T_b$ is relevant in some other "smoother" realization of $\mu$ or an artifact of the continuum approximation.

\no
The stability of the boundary states is determined by the stability matrix $\beta {\bf \Lambda C}$. For the standard continuum system
its eigenvalues and eigenstates satisfy
\eqn{stabex}, with the bound state configuration $x_b(s)$ instead of $x_k(s)$. Since $x_b(s)$ is nonzero only close to $s=0$,
by the same argument as for the modes $x_k$ the stability matrix is perturbatively close to the one of $x(s)=0$, that is
\be
{2j^2\over T} B_j' (0)\Big((c_0+2c) u(s)+ u''(s)\Big)= \l u(s)
\ee
and becomes identical to the one for the singlet. Since we are at $T>T_c$, the singlet is stable, and so is the boundary state.
We see that the magnet has two locally stable states for $T_b<T<T_B$, the boundary state and the singlet. Determining which one is absolutely stable
and which is metastable requires the evaluation and comparison of their free energies.

\subsubsection{Modified boundary conditions in the middle of the chain}

Changing the couplings between the spins in the neighborhood of any other point $s_0$ in the bulk of the chain would lead to conditions involving
$x(s_0)$ and the discontinuity of $x' (s)$ at $s=s_0$,
\be
\label{hhh}
\left. x'(s_0 +\varepsilon) - x'(s_0 -\varepsilon)\right|_{\varepsilon \to 0^+} = \mu x(s_0)\ ,
\ee
$\mu$ depending, in general, on the coupling constants and the temperature. This corresponds to a $\d$-function "kick" (impact) to the
fiducial particle at $s=s_0$ proportional to $x_{s_0}$.

\no
We compute $\mu$ in the situation where the single coupling between $x_0$ and $x_1$ is $c'$, all other couplings remaining equal to $c$. Working in the linearized
limit and setting $s_0 =0$ for convenience (with $s$ taking also negative values),
equations \eqn{xex} remain valid for all $s$ except $s=0,1$.
Unlike the case of the modified coupling at the end of the chain, however, these equations determine the values of all $x_s$, since they all participate in an
unmodified equation, but with the solution for $s\geqslant 1$ and $s\leqslant 0$ having, in general, different amplitudes and phases.
The equations for $s=0,1$ yield
\be
\label{c1c2n}
c x_{-1} + c' x_1 = 2c \cos\theta \, x_0\ ,\qq  c' x_0 + c x_2 = 2c \cos\theta \, x_1\ .
\ee
The equations for $x_s$ are invariant under the transformation $s \to 1-s$, so we write the solution for $x_s$ as a sum of an even and an odd part under
this transformation, each part satisfying independently the equations. (The two parts will be related by the Dirichlet boundary conditions at the ends of
the chain, so, in general, this transformation is not an invariance of the system unless the two ends are at equal distances from the pair $s=0,1$.)
The odd part of the solution, however, has vanishing $x(0)$ and a continuous $x'(0)$ in the continuum, and thus does not
contribute to $\mu$. We thus need to consider only the even part, which has the form
\be
x_s\sim \sin (\th |s-\half| + \d) \ .
\ee
Then \eqn{c1c2n} leads to the single equation
\be
{x_2\ov x_1}= {\sin(3 \th /2 +\d) \ov  \sin (\th /2+ \d)} 
= 2 \cos \th -{c'\ov c}\  ,
\ee
from which the phase shift is obtained as
\be
\cot\d = {c-c' \over c+c'} \cot{\th\over 2}\ .
\ee
Taking the position of the impurity to be at the point of symmetry of the solution, that is, $s=\half$,
the boundary condition becomes
\be
\mu = {2 x'(\half) \over x(\half)} = 2\theta \cot\d =  2 \th\, {c-c'\ov c+c'} \cot {\th\ov 2} \simeq 4{c-c'\ov c+c'}\ ,
\label{middle}\ee
The last expression is valid for $\th \ll 1$, which is the regime where the continuum limit becomes a good approximation.

\subsection{The periodic spin chain}

The case of periodic nearest-neighbor interactions, in which $x_m$ couples to $x_1$ with the same coupling constant $c$,
can be treated in a similar way and we simply state the results. 

\no
The system has now a discrete translation invariance $Z_m$, $x_s \to x_{s+n}$ ($n=0,1,\dots,m-1$), with periodic
boundary conditions $x_{s+m}=x_s$.
In the linearized approximation, states will be plane waves
\be
x_s = A \sin \th (s+s_0) = A_1 \cos \th s + A_2 \sin \th s \ , \quad \th = {2\pi k \over m} \ , \quad k=0,1,\dots, m-1 ,
\label{xsperiod}
\ee
The corresponding critical temperatures ${T}_k$ and eigenvalues ${\l}_k$ of 
the stability matrix $\b {\bf \Lambda} {\bf C}$ are
\be
{\l}_k = {{T}_k \over T} ~,\quad  T_k = {2j(j+1)\over 3T} \Big( c_0 + c \cos{2\pi k\over m}\Big) ~, ~~~k = 0,\dots,m-1
\label{lamm}\ee
The critical temperature above which the singlet is stable is $T_0 = T_c$, the same for all $m$.

\no
Since $T_k = T_{m-k}$, these temperatures are doubly degenerate and two independent states appear at each critical temperature,
with the exception of $k=0$, and $k=m/2$ for even $m$.
This degeneracy means that the thermodynamic states at each $T_k$ form a one-dimensional manifold,
parametrized by the continuous parameter $s_0$ in \eqn{xsperiod} (equivalently, by the ratio $A_1/A_2$), all states having the same free energy at $T= {T}_k$.
Below ${T}_k$, the magnetization increases (see \eqn{xsnonlin} in the open chain case) and the manifold of degenerate states reduces
to distinct states related by the discrete symmetry $Z_m$, each state breaking this symmetry.\footnote{For $k/m = p/q$, with $p,q$ relatively prime,
there are $q$ distinct states and $Z_m$ is broken to $Z_q$.} Transitions between states
require the system to go through configurations of higher free energy and are thermodynamically suppressed.
This suppression becomes energetically weaker but entropically stronger for large $m$, and transitions are exponentially suppressed
in the thermodynamic limit. 

\no
The $k=0$ constant state $x_s = \pm x_\text{max}$ for $T< T_0$, and (for even $m$ and $c<c_0$) the $k=m/2$ alternating state 
$x_s = \pm (-1)^s {\tilde x}_\text{max}$ for 
$T < {T}_{m\over 2}$ are $Z_m$-invariant and nondegenerate.
The amplitude of the constant mode $x_\text{max}$ satisfies \eqn{xm}, while the one of the alternating mode ${\tilde x}_\text{max}$, satisfies
\be
{\tilde x}_\text{max} = B_j \big(2j^2(c_0-c) {\tilde x}_\text{max}/T\big) \ ,\quad T<T_{m\over 2} = {2j(j+1) (c_0 -c) \over 3T} \ .
\ee
The stability of the constant mode is proved by eq.~\eqn{stamax}. For the alternating mode, the corresponding stability condition becomes
\be
\l_k = {2j^2\over T} B_j' \Big(2j^2 (c_0 -c) {\tilde x}_\text{max}/T \Big) \Big( c_0 - c \cos{2\pi k\over m}\Big) \leqslant 1
\, ,\quad k=0,1,\dots m-1 \ .
\ee
To hold for all values of $k$ it must hold for $k={m\over 2}$ that gives the highest $\l_k$, and the condition becomes
\be
{2j^2(c_0+c)\over T} B_j' \Big(2j^2 (c_0 - c) {\tilde x}_\text{max}/T \Big) \leqslant 1
\label{stacontil}\ee
A repetition of the argument that led to \eqn{stamax} now leads to the condition
\be
{2j^2(c_0+c)\over T} B_j' \Big(2j^2 (c_0 - c) {\tilde x}_\text{max}/T \Big) = 1 +{(2cT (\partial {\tilde x}_\text{max} / \partial T)+ (c_0 -c) {\tilde x}_\text{max}
\over (c_0 -c)( T (\partial {\tilde x}_\text{max} / \partial T) -x)}
\ee
The stability condition is not identically satisfied for temperatures below $T_{m\over 2}$. Since $\partial {\tilde x}_\text{max} / \partial T$ is negative, the denominator of
the fraction is negative, and to satisfy \eqn{stacontil} the numerator must be positive:
\be
2cT (\partial {\tilde x}_\text{max} / \partial T)+ (c_0 -c) {\tilde x}_\text{max} \geqslant 0
\ee
As $T\to T_{m\over 2}$ from below, ${\tilde x}_\text{max} \to 0$ and $\partial {\tilde x}_\text{max} / \partial T \to -\infty$, so the condition fails, while
for $T \to 0$, ${\tilde x}_\text{max} \to 1$ and $\partial {\tilde x}_\text{max} / \partial T \to 0$ and the condition is satisfied. There is, thus, some
temperature $0<{\tilde T}_c < T_{m\over 2}$ below which the alternating mode becomes stable, given by the solution of the transcendental
system of equations
\be
2j^2 (c_0 +c) B_j' (2j^2 (c_0-c) {\tilde x}_\text{max}/T) = 1 \ ,\quad B_j (2j^2 (c_0-c) {\tilde x}_\text{max}/T) = {\tilde x}_\text{max}
\ee
The remaining states for $T \lesssim T_k$, $0<k<{m\over 2}$, are unstable, since the stability
matrix $\b{\bf \Lambda C}$ for these states has (some) eigenvalues larger than 1.
As in the open chain case, these states become stable when the temperature drops substantially below $T_k$ and
the $x_s$ become large, as indicated by the continuum analysis.

\no
Note that $c \to -c$ is an automorphism of the system only for even $m$ (the corresponding redefinition $x_s \to (-1)^s x_s$ is
inconsistent with periodic boundary conditions for odd $m$) and maps $k \to {m\over 2} -k$. In particular, it maps the constant
mode to the alternating mode.

\no
In the continuum approximation, states for $T<T_c$ correspond to a full periodic motion of the fiducial particle such that $x(s) = x(s+m)$. States are similar in form
to those of the open chain but with an even number of half-periods and an arbitrary starting point $x(0)$, corresponding to the $Z_m \to U(1)$ symmetry of the chain
for $m\gg 1$. That is, they correspond to $x_{2k} (s-s_0)$ of the open chain modes with an arbitrary $s_0$. States become stable as they approach their
limiting form $x_{2k}(s-s_0) = \pm x_\text{max}$ away from the nodes at $s-s_0 ={nm\over 2k}$, $n=1,\dots 2k$.
Similar to the open chain case, for low $k$ these states become a train of $2k$ equidistant alternating kink-antikink solitons but with an 
arbitrary shift $s_0$. In addition,  for $T<T_0= T_c$ we have a stable constant (zero-soliton) mode $x(s) = \pm x_\text{max}$.
The alternating mode $x_s = (-1)^s {\tilde x}_\text{max}$ would correspond to a condensation of $m\over 2$ solitons alternating with $m\over 2$ antisolitons
and, as is oscillates at a scale $\sim1$, is not captured by the continuum approximation.

\no
Note that in the
continuum regime solitons always appear in pairs, in a soliton-antisoliton succession. This is enforced by the "topology" of the configuration space 
($x(s)$ has to pass through $x =0$ an even number of times and in opposite directions) and is reminiscent of similar restrictions in other physical 
one-dimensional systems.

\section{Concluding remarks }
\label{conclusions}

Linear magnetic macromolecules (magnetic chains) open a new range of possibilities for exotic magnetic behavior. The properties of the $SU(2)$
chain that we examined in this paper include a cascade of critical temperatures, metastable states, and the appearance of isolated "boundary" states coexisting with
the singlet above a critical temperature or with magnetized states below another critical temperature.  Interestingly, the states emerging at low
temperatures have the shape of isolated kink solitons of extent $\sim 1$ on a chain of large length $m$.

\no
In spite of the wealth of results and phenomena uncovered, it is clear that the analysis in this paper leaves many topics
open for future research. For one thing, we studied the emergence of magnetized states as solutions of the equilibrium equations.
These states are unstable when they appear, and become physical at some lower critical temperatures where they stabilize.
The calculation of these temperatures remains an
open issue, and, barring any analytical breakthrough, can only be achieved with numerical methods.

\no
An obvious direction for future investigation is the study of magnetic chains with atoms in representations of $SU(N)$, for $N>2$. Since even for a
single atom $SU(N)$ magnetism has qualitatively new features compared to those of standard $SU(2)$ magnets, we expect a greatly enlarged
set of phenomena to also appear in $SU(N)$ chains. It particular, it will be interesting to see if the classification of stable configurations in terms of Young
tableaux valid for the case of one atom \cite{PSferro,PShigher} persists or is modified.
In addition, the soliton states corresponding to magnetized states at low temperatures will now be "colored," with particular fusion rules when they coalesce.
The properties of soliton states, even in the simplest case of $SU(2)$ chains, is an exciting topic and the subject of ongoing work. 
Finally, the relevance of magnetic chains to real cold molecule systems \cite{Dud,Ghu,Gor,Zha,Mag,Cap,Mukherjee:2024ffz} and
to other theoretical models \cite{TLC,RQ,CFLT,RPAR},
and their possible connection to social interaction systems \cite{Nechaev}, are additional exciting topics for further investigation.

\subsection*{Acknowledgements }

 A.P.'s research was supported in part by the National Science Foundation 
under grant NFS-PHY-2112479, and by PSC-CUNY grants 69213-00 57 and 6D136-00 06.

\appendix

\section{Range of variation of $x(s)$ }

\label{stop}

\no
The question of the range of motion of the fiducial particle is interesting, and cannot simply be justified by the runaway behavior of the potential
$V(x)$ beyond $x=\pm x_\text{max}$, as the potential is finite at $x=\pm 1$ and the fate of the particle when it reaches these points is unclear.
To address it, we return to the discrete system and work separately for $c_0>0$ and $c_0<0$.

\no
For $c_0>0$, we work with the discrete equation \eqn{xsdis}. Since $x_s, x_{s\pm 1}$ on the left hand side are at most 1
and at least $-1$, we obtain
\ba
 {T\over j^2} G_j(|x_s |)\leqslant 2c_0+2c \quad \Longrightarrow &&|x_s |\leqslant B_j \Big({2j^2\over T} (c_0+ c)\Big)\, , \quad 1<s<m \ ,
 \label{xmax1} 
 \\
 {T\over j^2} G_j(|x_s |)\leqslant 2c_0+c \quad \Longrightarrow  &&|x_s |\leqslant B_j \Big({j^2\over T} (2c_0+ c)\Big)\, , \quad~ s=1,m\ .
  \label{xmax11}
\ea
So the fiducial particle never reaches $x=\pm1$. In fact, for spins not near the endpoints $s=1,m$, putting
the bound \eqn{xmax1} in \eqn{xsdis} and iterating the process, we obtain
\be
| x_s | \leqslant f\Big(f\big(f ( \dots (f(1)) \dots \Big)\, ,\quad \text{where} \quad f(x) = B_j \Big({2j^2\over T} (c_0+ c) x \Big)\ .
\ee
Since, for all $T$,
\be
\label{gfj1}
f( x_\text{max}) = x_\text{max} \quad \text{and}\quad  f' (x_\text{max}) <1\ ,
\ee
the above process converges to $| x_s | \leqslant x_\text{max}$ for $s$ far from $1$ or $m$.\footnote{\no
To prove $f'(x_\text{max}) <1$
note that $f(x)$ is concave for $x\geqslant 0$ and the mean value theorem implies
\be
\label{fpf}
f'(x_\text{max})< {f(x_\text{max})-f(0) \ov x_\text{max} - 0}={x_\text{max}-0 \ov x_\text{max} - 0}= 1 \ ,
\ee
If $f'(0)<1$ then $x_\text{max}=0$ and the inequality also holds.
}
For $s$ close to $1$ or $m$, in some iteration steps the inequality \eqn{xmax11} will be invoked.
However, since invoking \eqn{xmax11} gives a lower bound than \eqn{xmax1}, the final bound will be less than $x_\text{max}$. Therefore,
\be
|x_s | \leqslant x_\text{max} ~, \quad s = 1,\dots,m\ .
\ee

\no
For $c_0<0$, we write \eqn{xsdis} as
\be
{c} (x_{s-1} + x_{s+1}) =  {T \over j^2 }  G_j (x_s )-2c_0\, x_s \ .
\label{xsdis1}\ee
Since $x_{s\pm 1}$ on the left hand side are at most 1 and at least $-1$, we obtain
\ba
 -2c \leqslant {T\over j^2} G_j(x_s ) -2c_0  x_s \leqslant 2c \, , \qquad 1<s<m \ ,
 \label{xmax2}
  \\
 -c \leqslant {T\over j^2} G_j(x_s ) -2c_0  x_s \leqslant c \, , \qquad  ~~~~~ s=1,m\ .
\label{xmax22}\ea
The function of $x_s$ appearing above is odd and, for $c_0<0$, monotonically increasing. Calling its inverse $H(\, \cdot \,)$,
\eqn{xmax2} and \eqn{xmax22} become
\ba
{T\over j^2} G_j(|x_s| ) -2c_0  |x_s| \leqslant 2c \, \quad \Longrightarrow &&|x_s |\leqslant H(2c)\, ,\qquad 1<s<m \ ,
\label{xmax3}
 \\
{T\over j^2} G_j(|x_s| ) -2c_0  |x_s| \leqslant c \, \qquad \Longrightarrow &&|x_s |\leqslant H(c)\, , ~~~~~ s=1,m\ .
\label{xmax33}
\ea
As before, for spins not near the endpoints $s=1,m$, putting
the bound \eqn{xmax3} in \eqn{xsdis1} and iterating the process, we obtain
\be
| x_s | \leqslant F\Big(F\big(F ( \dots (F(1)) \dots \Big)\, ,\quad \text{where} \quad F(x) = H \big(2cx\big)\ .
\ee
Since, for all $T$,
\be
\label{hhh1}
F( x_\text{max}) = x_\text{max} \quad \text{and}\quad  F' (x_\text{max}) <1\ ,
\ee
the above process again converges to $| x_s | \leqslant x_\text{max}$ for $s$ far from $1$ or $m$.\footnote{Note that $F(x)$ is concave, since it is the inverse
of the convex function $ {T\ov j^2} G_j(2 c x) - 2 c_0 c x$, so the mean value theorem implies $F'(x_\text{max})<1$ as in the case $c_0>0$.}
For $s$ close to $1$ or $m$, in some iteration steps the inequality \eqn{xmax33} will be invoked, so the process will converge to a different value.
However, since invoking \eqn{xmax33} always gives a lower bound, the final bound will be less than $x_\text{max}$. Therefore, for any $c_0$,
\be
|x_s | \leqslant x_\text{max} ~, \quad s = 1,\dots,m\ .
\label{xsbound}
\ee
In particular, for $T>T_c$, $x_\text{max} = 0$ and we obtain $| x_s | \leqslant 0$, which shows that $x_s =0$ is the only solution.

\no
The bound \eqn{xsbound} can be violated only if some coupling constants are different than $c_0,c$, in which case some of the inequalities
in \eqn{xmax1} are modified. This is relevant to the case of modified boundary conditions for the chain, where the coupling $C_{1,2}$ is different than $c$. 
In this case, the inequality for $s=1$ becomes (for $c_0>0$)
\be
{T\over j^2} G_j(|x_1 |)\leqslant 2c_0+c'\ .
\ee
For this inequality to be less restrictive than the one for the other sites, thus allowing a possible upper bound higher than $x_\text{max}$, we
must have $c' > 2c$. 

\section{Boundary state analysis in the discrete system}\label{fullmonty}

We have studied the emergence of boundary states in two regimes, the linearized limit and the continuum limit. Both are approximations, working
for small magnetizations or for low chain modes (slow change of magnetization with $s$). The linearized approximation can, in fact, reliably identify
the onset of boundary modes at some critical temperatures, but the persistence of these modes for temperatures below the critical ones is uncertain.

\no
To gain confidence on the existence and persistence of boundary states, we return to the discrete system beyond the linear
approximation. We will focus, for concreteness, on boundary states that do not oscillate rapidly with $s$; that is, states corresponding to $T\simeq T_B$
in the linearized analysis (top line of \eqn{xbound}). As before, all couplings are equal to $c$ except the one between $x_1$ and $x_2$ which is $c'$.

\no
The appearance of such boundary states can be understood as follows: we consider a solution of the equilibrium equations for which $x_s$ is small
over values $s$ far away from $s=1$ and decays exponentially with $s$. For $T>T_c$, the decay rate can be calculated from equation
\eqn{xlin}, valid for small $x_s$, giving the state as
\be
x_s = A e^{-\ka s} ~,\quad \cosh \ka = 1+ {3(T-T_c) \over 2j(j+1)c}\ .
\label{oura}\ee
As $s$ decreases, $x_s$ will become large and the exact discrete equations have to be used. We write them as
\be
x_{s-1} = -x_{s+1} -{2c_0 \over c} x_s + {T\over j^2 c} G_j (x_s)\ ,
\label{sdown}\ee
valid for $s\geqslant 3$, where the couplings of $x_s$ to $s+1$ and $s-1$ are both equal to $c$. The remaining two equations, written in a suggestive form, are
\be
{c' \over c} x_1 = - x_3 - 2{c_0\over c} x_2 + {T\over cj^2} G_j (x_2) ~, \qq  0 = -{c'\over c} x_2 - 2{c_0\over c} x_1 + {T \over cj^2} G_j (x_1)\ .
\label{s12}
\ee
If $c'=c$, the equation for $x_1$ is part of the set of equations \eqn{sdown}, while the last equation can be viewed as
calculating $x_0$ from \eqn{sdown} and demanding it to be zero.

\no
Equation \eqn{sdown} can be treated recursively, producing $x_{s-1}$ from $x_{s+1}$ and $x_s$.
Starting with \eqn{oura} and working towards smaller $s$, the nonlinearity of \eqn{sdown} will eventually become important. Assuming $c_0,c>0$
and $x_s >0$, we can show that the sequence will be increasing as $s$ decreases with a rate at least as large as $e^\ka$. Indeed, given that
\be
G_j (x) > G'_j(0) x = {3j \over j+1} x\ ,
\label{Gbig}\ee
for $x>0$, and assuming $x_s > e^\ka x_{s+1}$ for some $s$, we have from \eqn{sdown}
\ba
x_{s-1} \bb&>&\bb - e^{-\ka} x_s - {2c_0 \over c} x_s +{3T \over j(j+1)c} x_s \cr
&=&\Bigl( -e^{-\ka} + {3(T-T_c)  \over j(j+1) c} +2 \Bigr) x_s\cr
&=& e^\ka x_s\ ,
\ea
where in the last step we used the equation for $\ka$ \eqn{oura}. Thus, by induction, $x_{s-1} > e^\ka x_s$ for all smaller $s$.
This result can be used to prove that the singlet $x_s =0$ is the only allowed state in the undeformed chain: as we stated,
the second equation in \eqn{s12} sets $x_0=0$, which is not possible since, as shown, $x_0 > e^\ka x_1$. 

\no
The sequence \eqn{sdown} will stop by itself whenever the calculated value of $x_{s-1}$
exceeds $1$ for some $s$, which is an unphysical value. If the process persists all the way to $s=3$, it will still need to satisfy \eqn{s12}.
For $c' \neq c$, the $x_s$ down to $s=2$ are those produced by \eqn{sdown}. In the last two equations
\eqn{s12} we can view both $x_1, x_2$ as free variables: $x_1$ by default, and $x_2$ by 
adjusting the scale $A$ of the exponentially decaying part of the sequence \eqn{oura} (this also
fixes $x_3$). If a solution  $x_1,x_2$ of \eqn{s12} exists, it will represent a boundary state.

\no
We start by giving a proof of existence of boundary states, by inverting the problem: fixing $x_2$ and adjusting $x_1$ and $c'$ to satisfy \eqn{s12}.
The first equation in \eqn{s12} is the same as \eqn{sdown} for $s=2$, with ${c' x_1 /c}$ playing the role of $x_1$. Calling ${\bar x}_1$ the value of $x_1$ produced
by \eqn{sdown}, we have
\be
c' = {c {\bar x}_1 \over x_1}
\label{fixc}\ee
and substituting this expression in the second equation in \eqn{s12} we obtain
\be
c {\bar x}_1 x_2 = - 2c_0\,\,\bb x_1^2 + {T \over j^2}\, x_1 G_j (x_1)\ . 
\label{cfromx}
\ee
The function of $x_1$ in the right hand side starts from $0$ at $x_1 =0$ and monotonically increases to infinity at $x_1 =1$. Therefore,
this equation always has a unique solution for $x_1<1$, which also fixes $c'$ from \eqn{fixc}. This proves the existence of boundary states for
a specific $c'$ depending on $T$ and the choice of $x_2$. Varying $x_2$ from $0$ to $1$, and thus ${\bar x}_1$ from $0$ to infinity, we can in principle
match the so-obtained value of $c'$ with the actual coupling $c'$ of the system.

\no
It remains to see if this process will produce all possible values of $c'$, or, stated differently, for what values of temperature will boundary states
exist for a given $c'$. To do that, we examine the constraints that we have already derived. Since ${\bar x}_1= c' x_1 /c$ is the quantity derived from
application of \eqn{sdown}, it obeys the constraint ${\bar x}_1 > e^\ka x_2$, which leads to
\be
x_2 < e^{-\ka}\,{c' x_1 \over c}\ .
\ee
Applying this inequality, as well as \eqn{Gbig}, in the second equation \eqn{s12}, we obtain
\ba
e^{-\ka} {c'^2\over c} x_1 &>& {3T \over j(j+1)} x_1 -c_0 x_1 = 2c \Bigl({3(T-T_c)\over 2j (j+1)c} +1\Bigr) x_1
= c\big(e^\ka + e^{-\ka}\big) x_1
\ea
where we used again the equation for $\ka$. Altogether,
\be
\Big( {c'^2 \over c^2} -e^{2\ka} -1 \Big) x_1 > 0\ .
\label{cbig}\ee
Since $x_1>0$, its coefficient must be positive. This means that $c'$ cannot be smaller than a critical value saturating the inequality
at a given temperature $T$, or, conversely, that there is a critical temperature $T_B$ above which the boundary state does not exist.
Using the expression for $\k$ in \eqn{cbig} we obtain
\be
T < T_B = T_c + {j(j+1)\over 3} \Bigl({c'^2 \over \sqrt{c'^2 - c^2}}-2c\Bigr) = {j(j+1)\over 3} \Big(2c_0 +{c'^2 \over \sqrt{c'^2 - c^2}}\Big)\ .
\label{rea}
\ee
This is the same as $T_B$ in \eqn{TBTL} found in the linear regime.
Indeed, the inequality can be saturated assuming small values of $x_1,x_2$. For $T$ smaller but very close to $T_B$, we set
\be
T = T_B (1-\epsilon) ~,\quad \epsilon \ll 1
\ee
and take $x_1,x_2 \ll 1$. In this case, we are still in the linear regime \eqn{oura}
and ${\bar x}_1 = e^\ka x_2$.
Then, $x_1 = c {\bar x}_1 /c' = c e^\ka x_2 /c'$. Putting $x_2 = c' e^{-\ka} x_1 /c$ in the second equation \eqn{s12}, and expanding
\be
G_j (x) = {3j\over j+1} x + {9j(2j^2 +2j+1)\over 10 
(j+1)^3} x^3 + \cdots\ ,
\ee
we obtain 
\be
e^{-\ka}\Big({c'^2\over c^2}-e^{2\ka} -1\Big) x_1 = {3(2j^2+2j+1) T \over 10 j(j+1)^3 c} x_1^3 + \cdots\ .
\ee
Putting $T=T_B(1-\epsilon)$ and using the expressions for $\ka$ and $T_B$, we obtain, to order $\epsilon$,
\be
x_1 = 4c(j+1)\sqrt{ {5 j\, (T_B - T)  \over 3(2j^2+2j+1) (c'^2 - c^2) T_B}}\ .
\ee
Thus, close to $T_B$ the amplitude of the bounday state behaves
as $(T_B -T)^{1/2}$, as is typical of phase transitions, and at $T=T_B$ it vanishes, leaving the singlet as the only state.
Note that, in order for boundary states to exist at all, the condition $c'^2 > 2 c^2$ must hold, marking the point at which $T_B = T_c$,
consistent with \eqn{cpcr} for $m\gg 1$.

\no
We established that $T_B$ is the maximum temperature for which a bound state exists. On the other hand, we found no minimum temperature
for the formation of bound states other than $T_c$. In fact, \eqn{cfromx}
shows that we can obtain an arbitrarily high $c'$ at any fixed $T>T_c$: by increasing the amplitude of the boundary state $A$ we can bring $x_2$
close to $1$, driving ${\bar x}_1$ to arbitrarily large values. Equation \eqn{cfromx} will have a solution for $x_1$ close to $1$, leading to arbitrarily
large values of $c'$ through \eqn{fixc}. This means that there is no maximum value of $c'$ at any $T$, or, conversely, that there is no minimum temperature
$T_b$ for fixed $c'$ below which the boundary state ceases to exist.

\no
In conclusion, the temperature $T_b$ in \eqn{Bcon} of the continuum limit does not appear in the realization of general boundary conditions with only
one coupling between $x_1$ and $x_2$ different from $c$ in the discrete system. It could still be that such a temperature does
exist in situations where several couplings near $s=1$ are different from $c$, realizing generalized boundary conditions in a "smoother" way that is better
captured by the continuum limit. This possibility remains, as yet, unexplored.


\begin{thebibliography}{99.}


\bibitem{Dud}
R.A.~Duine and A.H.~MacDonald,
{\it Itinerant Ferromagnetism in an Ultracold Atom Fermi Gas},
\href{https://link.aps.org/doi/10.1103/PhysRevLett.95.230403}{Phys. Rev. Lett. \textbf{95} (2005) 230403}

\bibitem{Ghu}
M.A. Cazalilla, A.F. Ho and M. Ueda,
{\it Ultracold gases of ytterbium: ferromagnetism and Mott states in an $SU(6)$ Fermi system},
\href{https://iopscience.iop.org/article/10.1088/1367-2630/11/10/103033}{New J. Phys. {\bf 11}
(2009) 103033}.

\bibitem{Gor}
A.V.~Gorshkov {\it et al.}, {\it Two-orbital SU(N) magnetism with ultracold alkaline-earth atoms},
\href{https://www.nature.com/articles/nphys1535}{Nature Physics \textbf{6} (2010) 289}.

\bibitem{Zha}
X. Zhang  {\it et al.}, {\it Spectroscopic observation of $SU(N)$-symmetric interactions in Sr orbital
magnetism}, \href{https://www.science.org/doi/abs/10.1126/science.1254978}
{Science {\bf 345}, Issue 6203 (2014) 1467}.

\bibitem{Mag}
M.A. Cazalilla and A.M. Rey, {\it Ultracold Fermi gases with emergent $SU(N)$ symmetry}, 
\href{https://iopscience.iop.org/article/10.1088/0034-4885/77/12/124401}{Rep. Prog. Phys. {\bf 77} (2014) 124401}.

\bibitem{Cap}
S.~Capponi, P.~Lecheminant, and K.~Totsuka, {\it Phases of one-dimensional SU(N) cold atomic 
Fermi gases - From molecular Luttinger liquids to topological phases}, \hfill\break 
\href{https://www.sciencedirect.com/science/article/abs/pii/S0003491616000130}{Ann. Phys. \textbf{367} (2016) 50}.

\bibitem{Mukherjee:2024ffz}
B.~Mukherjee, J.M.~Hutson and K.~R.~A.~Hazzard,
{\it $SU(N)$ magnetism with ultracold molecules}, New J. Phys. {\bf 27} (2025) 013013,
\href{https://arxiv.org/abs/arXiv:2404.15957}{arXiv:2404.15957 [cond-mat]}.

\bibitem{Aff}
I.~Affleck, {\it Large-$n$ Limit of $SU(n)$ Quantum "Spin" Chains},\hfill\break
\href{https://journals.aps.org/prl/abstract/10.1103/PhysRevLett.54.966}{Phys. Rev. Lett. \textbf{54} (1985) 966}.

\bibitem{Pola}
A.P.~Polychronakos, {\it Exact Spectrum of $SU(n)$ Spin Chain with Inverse-Square Exchange},
Nucl. Phys. \textbf{B419} (1994) 553,
\href{http://arxiv.org/abs/hep-th/9310095}{arXiv:9310095 [hep-th]}.


\bibitem{KT}
H. Katsura and A. Tanaka,
{\it Nagaoka states in the $SU(n)$ Hubbard model}, \hfill\break
\href{https://journals.aps.org/pra/abstract/10.1103/PhysRevA.87.013617}{Phys. Rev. {\bf A87}  (2013) 013617}.

\bibitem{BSL}
E. Bobrow, K. Stubis and Y. Li,
{\it Exact results on itinerant ferromagnetism and the 15-puzzle problem},
\href{https://journals.aps.org/prb/abstract/10.1103/PhysRevB.98.180101}
{Phys. Rev. {\bf B98}  (2018) 180101(R)}.

\bibitem{RoLa}
C. Romen and A.M. L\"auchli, {\it Structure of spin correlations in high-temperature $SU(N)$
quantum magnets},
 \href{https://journals.aps.org/prresearch/abstract/10.1103/PhysRevResearch.2.043009}{Phys. Rev. Research {\bf 2}
(2020) 043009}.

\bibitem{YSMOF}
D. Yamamoto, C. Suzuki, G. Marmorini, S. Okazaki and N. Furukawa,
{\it Quantum and Thermal Phase Transitions of the Triangular $SU(3)$ Heisenberg Model under Magnetic Fields},
\href{https://journals.aps.org/prl/abstract/10.1103/PhysRevLett.125.057204} {Phys. Rev. Lett. {\bf 125} (2020) 057204} .

\bibitem{TK}
 K. Tamura and H. Katsura, 
{\it   Ferromagnetism in $d$-Dimensional $SU(n)$ Hubbard Models with Nearly Flat Bands},
\href{https://link.springer.com/article/10.1007/s10955-020-02687-w}{Journal of Stat. Phys. {\bf 182} (2021) 16}.

\bibitem{Totsuka} K. Totsuka, 
{\it Ferromagnetism in the $SU(N)$ Kondo lattice model: $SU(N)$ double exchange and supersymmetry},
\href{https://journals.aps.org/pra/abstract/10.1103/PhysRevA.107.033317} {Phys. Rev. {\bf A107} (2023) 033317}.

\bibitem{TK2}
K. Tamura, H. Katsura
{\it Flat-band ferromagnetism in the SU(N) Hubbard and Kondo lattice models},
\href{https://arxiv.org/abs/2303.15820}{arXiv:2303.15820 [cond-mat]}.

\bibitem{DY}
D.~Yamamoto {\it et al.}, {\it Quantum and Thermal Phase Transitions of the Triangular $SU(3)$ Heisenberg Model under Magnetic Fields}, 
\hfill\break 
Phys. Rev. Lett. \textbf{125} (2020) 057204,
\href{https://arxiv.org/abs/2004.03250}{arXiv:2004.03250 [cond-mat]}.

\bibitem{YM}
Y.~Miyazaki {\it et al.},
{\it Linear Flavor-Wave Analysis of $SU(4)$-Symmetric Tetramer Model with Population Imbalance},\hfill\break
J. Phys. Soc. Jpn. {\bf 91} (2022) 073702, 
\href{https://arxiv.org/abs/2205.11155}{arXiv:2205.11155 [cond-mat]}.

\bibitem{HM}
H. Motegi. G. Marmorini, N. Furukawa, D. Yamamoto, {\it Thermal Ising transition in
two-dimensional $SU(3)$ Fermi lattice gases with population imbalance}, \hfill\break
Phys. Rev. Research {\bf 5} (2023) L022056,
\href{https://arxiv.org/abs/2209.05919}{arXiv:2209.05919 [cond-mat]}.

\bibitem{PSferro}
A.P.~Polychronakos and K.~Sfetsos,
{\it Ferromagnetic phase transitions in SU(N)}, \hfill\break
Nucl. Phys. \textbf{B996} (2023) 116353,
\href{https://arxiv.org/abs/2306.01051}{arXiv:2306.01051 [hep-th]}.


\bibitem{PSlargeN}
A.P.~Polychronakos and K.~Sfetsos,
{\it Phase transitions in the decomposition of $SU(N)$ representations},
Nucl. Phys. \textbf{B999} (2024) 116434,
\href{https://arxiv.org/abs/2310.16887}{arXiv:2310.16887 [hep-th]}.


\bibitem{PSferroN}
A.P.~Polychronakos and K.~Sfetsos,
{\it Triple critical point and emerging temperature scales in SU(N) ferromagnetism at large N}, \hfill\break
Nucl. Phys. \textbf{B1009} (2024) 116748,
\href{https://arxiv.org/abs/2408.08357}{arXiv:2408.08357 [hep-th]}.


\bibitem{PShigher}
A.~P.~Polychronakos and K.~Sfetsos,
{\it Ferromagnets from higher SU(N) representations},
Nucl. Phys. \textbf{B1014} (2025) 116880,
\href{https://arxiv.org/abs/arXiv:2501.11681}{arXiv:2501.11681 [hep-th]}.

\bibitem{PScubic}
A.P.~Polychronakos and K.~Sfetsos,
{\it Nonabelian ferromagnets with three-body interactions},
Nucl. Phys.  \textbf{B1018} (2025) 117094,
\href{https://arxiv.org/abs/arXiv:2506.19916}{arXiv:2506.19916 [hep-th]}.

\bibitem{LPSadj}
J.~L{\'o}pez-Su{\'a}rez, A.~P.~Polychronakos and K.~Sfetsos,
{\it Adjoint ferromagnets},
Nucl. Phys. \textbf{B1023} (2026), 117287,
\href{https://arxiv.org/abs/arXiv:2510.11776}{arXiv:2510.11776 [cond-mat.stat-mech]}.

\bibitem{McCoy-Wu}
B.M. McCoy and  T.T. Wu, {\it Theory of a Two-Dimensional Ising Model with Random Impurities. I. Thermodynamics}, 
\href{https://doi.org/10.1103/PhysRev.176.631}{Phys. Rev. {\bf 176} (1968) 631};
{\it Theory of a Two-Dimensional Ising Model with Random Impurities. II. Spin Correlation Functions},  \href{https://doi.org/10.1103/PhysRev.188.982}{ Phys. Rev. {\bf 188} (1969) 982}.

\bibitem{McCoy}
B.M. McCoy,
{\it Theory of a Two-Dimensional Ising Model with Random Impurities. III. Boundary Effects}, \href{https://doi.org/10.1103/PhysRev.188.1014}{Phys. Rev. 
{\bf 188} (1969) 1014}. 

\bibitem{McCoyMean}
B. Berche, P.E. Berche, F. Igloi, G.  Palagyi. {\it The McCoy-Wu Model in the Mean-field Approximation}, J. Phys. {\bf A31} (1998) 5193,
\href{https://arxiv.org/abs/cond-mat/9804132}{arXiv:cond-mat/9804132 [cond-mat.stat-mech]}.

 
\bibitem{mols}
A.~P.~Polychronakos and K.~Sfetsos,
{\it Interacting multicomponent $SU(N)$ magnets}, 
\href{https://arxiv.org/abs/arXiv:2610.07157}{arXiv:2610.07157 [hep-th]}.

\bibitem{PScompo}
A.P.~Polychronakos and K.~Sfetsos,
{\it Composing arbitrarily many $SU(N)$ fundamentals},
Nucl. Phys.  \textbf{B994} (2023) 116314,
\href{https://arxiv.org/abs/arXiv:2305.19345}{arXiv:2305.19345 [hep-th]}.


\bibitem{TLC}
K. Totsuka, P. Lecheminant, and S. Capponi,
{\it Semiclassical approach to competing orders in a two-leg spin ladder with ring exchange},
\href{https://journals.aps.org/prb/abstract/10.1103/PhysRevB.86.014435}{Phys. Rev. {\bf B86} (2012) 014435}.

\bibitem{RQ}
A. Roy and T. Quella
{\it Chiral Haldane phases of $SU(N)$ quantum spin chains},\hfill\break
\href{https://journals.aps.org/prb/abstract/10.1103/PhysRevB.97.155148}{Phys. Rev. {\bf B97} (2018) 155148},
\href{https://arxiv.org/abs/1512.05229}{arXiv:1512.05229[cond-mat.str-el]}.


\bibitem{CFLT}
S. Capponi, P. Fromholz, P. Lecheminant and K. Totsuka,
{\it Symmetry-protected topological phases in a two-leg SU(N) spin ladder with unequal spins},\hfill\break
\href{https://journals.aps.org/prb/abstract/10.1103/PhysRevB.101.195121}{Phys. Rev. {\bf B101} (2020)
195121}, \href{https://arxiv.org/abs/1909.13553}{arXiv:1909.13553 [cond-mat.str-el]}.

\bibitem{RPAR}
K. Remund, R. Pohle, Y. Akagi, J. Romhanyi and N. Shannon,
{\it Semi-classical simulation of spin-1 magnets},
\href{https://journals.aps.org/prresearch/abstract/10.1103/PhysRevResearch.4.033106} 
{Phys. Rev. Res. {\bf 4} (2022) 033106}.


\bibitem{Nechaev}
P. Akarapipattana, S. Nechaev and B. Slavov,
{\it Collective communication in a transparent world: Phase transitions in a many-body Potts model and social-quantum duality},
 Phys. Rev.  {\bf E114} (2026) 024107,
\href{https://arxiv.org/abs/arXiv:2508.20267}{arXiv:2508.20267 [hep-th]}.


\end{thebibliography}
\end{document}